\documentclass[vecphys]{svmult}

\usepackage{graphicx}
\usepackage{amsmath,amssymb}
\usepackage{pstcol}
\usepackage{pspicture}
\usepackage{pst-plot}
\usepackage[bottom]{footmisc}
\usepackage[square,comma,sort&compress]{natbib}

\def\cs#1#2{#1_{\!{}_#2}}
\def\css#1#2#3{#1^{#2}_{\!{}_#3}}
\def\ket#1{|#1\rangle}

\begin{document}

\title*{Absolute Being vs Relative Becoming\thanks{To appear in
{\it Relativity and the Dimensionality of the World} (within the series {\it
Fundamental Theories of Physics}), edited by Vesselin Petkov (Springer, NY 2007).}}

\author{Joy Christian\inst{1,}\inst{2}}

\institute{Perimeter Institute, 31 Caroline Street North, Waterloo, ON, N2L 2Y5, Canada
\texttt{jchristian@perimeterinstitute.ca}
\and Department of Physics, Oxford University, Oxford OX1 3PU, United Kingdom
\texttt{joy.christian@wolfson.ox.ac.uk}}

\maketitle

\begin{abstract}
Contrary to our immediate and vivid sensation of past, present, and future as continually
shifting non-relational modalities, time remains as tenseless and relational as space in all
of the established theories of fundamental physics. Here an empirically adequate generalized
theory of the inertial structure is discussed in which proper time is causally compelled to
be {\it tensed} within both spacetime and dynamics. This is accomplished by introducing the
inverse of the Planck time at the conjunction of special relativity and Hamiltonian mechanics,
which necessitates energies and momenta to be invariantly bounded from above, and lengths and
durations similarly bounded from below, by their respective Planck scale values. The resulting
theory abhors any form of preferred structure, and yet captures the transience of {\it now}
along timelike worldlines by causally necessitating a genuinely becoming universe. This is
quite unlike the scenario in Minkowski spacetime, which is prone to a block universe
interpretation. The minute deviations from the special relativistic effects such as dispersion
relations and Doppler shifts predicted by the generalized theory remain quadratically suppressed
by the Planck energy, but may nevertheless be testable in the near future, for example via
observations of oscillating flavor ratios of ultrahigh energy cosmic neutrinos, or of altering
pulse rates of extreme energy binary pulsars.
\end{abstract}

\newenvironment{Quote}
  {\begin{list}{}{%
      \setlength{\rightmargin=0.5cm}{\leftmargin=0.5cm}}
        \item[]\ignorespaces}
  {\unskip\end{list}}

\parindent 0.5cm

\section{Introduction}\label{sec:1}

From the very first imprints of awareness, ``change'' and ``becoming'' appear to us to be two
indispensable norms of the world. Indeed, {\it prima facie} it seems impossible to make sense
of the world other than in terms of changing things and happening events through the incessant
passage of time. And yet, the Eleatics, led by Parmenides, forcefully argued that change is
nothing but an illusion, thereby rejecting the prevalent view, expounded by Heraclitus, that
becoming is all there is. The great polemic that has ensued over these two diametrically
opposing views of the world has ever since both dominated and shaped the course of western
philosophy \cite{Popper-1998}. In modern times, influential neo-Eleatics such as McTaggart
have sharpened the choice between the being and the becoming universe by distinguishing two
different possible modes of temporal discourse, one with and the other without a clear
reference to the distinctions of {\it past}, {\it present}, and {\it future}; and it is the
former mode with explicit reference to the tenses that is deemed essential for capturing the
notions of change and becoming \cite{McTaggart-1908}. Conversely, the latter mode---which
relies on a {\it tenseless} linear ordering of temporal moments by a transitive,
asymmetric, and irreflexive relation {\it precedes}---is deemed incapable of describing a
genuine change or becoming. Such a sharpening of the temporal discourse, in turn, has inspired
two rival philosophies of time, each catering to one of the two possible modes of the
discourse \cite{Smith-1994}. One
{\it tenseless} philosophy of time holds that time is {\it relational}, much like space, which
clearly does not seem to ``flow'', and hence what we perceive as the flow or passage of time
must be an illusion. The other {\it tensed} philosophy of time holds, on the other hand, that
there is more to time than mere relational ordering of moments. It maintains that time is rather
a dynamic or evolving entity {\it unlike} space, and does indeed ``flow''---like a
refreshing river---much in line with our immediate experience of it. That is to say, far from
being an illusion, our sensation of that sumptuous moment {\it now}, ceaselessly streaming-in
from nowhere and slipping away into the unchanging past, happens to reflect a truly objective
feature of the world.

In terms of these two rival philosophies of time, a genuinely becoming universe must then
correspond to a notion of time that is more than a mere set of ``static'' moments, linearly
ordered by the relation {\it precedes}. In addition, it must at least allow a genuine partition
of this ordered set into the moments of {\it past}, {\it present}, and {\it future}. From the
perspective of physics, the choice of a becoming universe must then necessitate a theory of
space and time that not only distinguishes the future events from the past ones intrinsically,
but also thereby accounts for the continual passage of the {\it fleeting present}, from a
{\it non-existing future} into the {\it unalterable past}, as a {\it bona fide} structural
attribute of the world. Such a theory of space and time, which would account for the gradual
{\it coming-into-being} of the non-existent future events---or a continual accumulation of the
unalterable past ones---giving rise to a truly becoming universe, may be referred to as a
{\it Heraclitean} theory of space-time, as opposed to a {\it Parmenidean} one, devoid of any
such explicit dictate to becoming.

One such Heraclitean theory of space-time was, of course, that of Newton, for whom
``[a]bsolute, true, and mathematical time, of itself, and from its own nature, flow[ed]
equably without relation to anything external...'' \cite{Newton-1671}. To be sure, Newton well
appreciated the relational attributes of time, and in particular their remarkable
similarities with those of space:
\begin{Quote}
Just as the parts of duration are individuated by their order, so that (for example) if
yesterday could change places with today and become the later of the two, it would lose its
individuality and would no longer be yesterday, but today; so the parts of space are
individuated by their positions, so that if any two could exchange their positions, they would
also exchange their identities, and would be converted into each other {\it qua} individuals.
It is only through their reciprocal order and positions that the parts of duration and space
are understood to be the very ones that they truly are; and they do not have any other principle
of individuation besides this order and position \cite{Hall and Hall-1962}.
\end{Quote}
And yet, Newton did not fail to recognize the {\it non-relational}, or absolute, attributes of
time that go beyond the mere relational ordering of moments. He clearly distinguished his
neo-platonic notion
of ``equably'' flowing absolute time, existing independently of changing things, from the
Aristotelian notion of ``unequably'' flowing relative times, determined by their less than
perfect empirical measures (such as clocks) \cite{Newton-1671}. What is more, he well
appreciated the closely related need of a temporally founded theory of calculus within
mathematics, formulated in terms of his notion of {\it fluxions} (i.e., continuously generated
temporally flowing quantities \cite{Arthur-1995}), and defended this theory vigorously against
the challenges that arose from the quiescent theory of calculus put forward
by Leibniz \cite{Arthur-1995}. Thus, the notions of flowing time and becoming universe were
central to Newton not only for his mechanics, but also for his mathematics \cite{Arthur-1995}.
More relevantly for our purposes, according to him the {\it rate} of flow of time---i.e., the
{\it rate} at which the relationally ordered events succeed each other in the world---is
determined by the respective moments of his absolute time, which flows by itself, continuously,
uniformly, and unstoppably, without relation to anything external \cite{Whitrow-1980}. Alas, as
we now well know, such a Newtonian theory of externally flowing absolute time, giving rise to
an objectively becoming universe, is no longer physically viable. But is our celebration of
Einstein's relativistic revolution complete {\it only} through an unconditional
renunciation of Newton's non-relationally becoming universe?

The purpose of this essay, first, is to disentangle the notion of a becoming universe from that
of an absolute time, and then to differentiate two physically viable and empirically
distinguishable theories of spacetime: namely, special relativity---which is prone to a
{\it Parmenidean} interpretation---and a generalized theory \cite{Christian-2004}---which
is intrinsically {\it Heraclitean} by construction. The purpose of this essay may also be taken
as a case study in {\it experimental metaphysics}, since it evaluates conceivable experiments
that can adjudicate between the two rival philosophies of time under discussion. Experimental
metaphysics is a term suggested by Shimony \cite{Shimony-1993a} to describe the enterprise of
sharpening of the disputes traditionally classified as metaphysical, to the extent that they
can be subjected to controlled experimental investigations. A prime example of such an
enterprise is the sharpening of a dispute over the novel conceptual implications of quantum
mechanics, which eventually led to a point where empirical evidence was brought to bear
on the traditionally metaphysical concerns of scientific realism \cite{Shimony-1993a}.
Historically, recall how resistance to accept the novel implications of quantum mechanics had
led to suggestions of alternative theories---namely, hidden variable theories. Subsequently, the
efforts by de Broglie, von Neumann, Einstein, Bohr, Bohm and others led to theoretical sharpening
of the central concepts of quantum mechanics, which eventually culminated into Bell's incisive
derivation of his inequalities. The latter, of course, was a breakthrough that made it possible
to experimentally test the rival metaphysical positions on quantum mechanics
\cite{Shimony-1993b}.
As this well known example indicates, however, experimental investigations alone cannot
be expected to resolve profound metaphysical questions once and for all, without careful
conceptual analyses. Indeed, Shimony \cite{Shimony-1993a} warns us against overplaying
the significance of experimental metaphysics. He points out that without careful conceptual
analyses even those questions that are traditionally classified as scientific cannot be resolved
by experimental tests alone. Hence, it should not be surprising that questions as slippery as
those concerning time and becoming would require more than a mere experimental input. On the
other hand, as the above example proves, a judicious experimental input {\it can, indeed},
facilitate greatly towards a possible resolution of these questions.

Bearing these cautionary remarks in mind, the question answered, affirmatively, in the present
essay is: Can the debate over the being vs becoming universe---which is usually also viewed as
metaphysical \cite{Schlesinger}---be sharpened enough to bear empirical input? Of course, as the
above example of hidden variable theories suggests, the first step towards any empirical effort
in this direction should be to construct a physically viable Heraclitean alternative to special
relativity. As alluded to above, this step has already been taken in Ref. \cite{Christian-2004},
with motivations for it stemming largely from the temporal concerns in quantum gravity. What is
followed up here is a comparison of these two alternative theories of causal structure with
regard to the status of becoming. Accordingly, in the next section we begin by reviewing the
status of becoming within special relativity. Then, in Sec. \ref{sec:3}, we review the
alternative to special relativity proposed in Ref. \cite{Christian-2004}, with an emphasis in
Subsec. \ref{subsec:3.3} on the causal inevitability of the strictly Heraclitean character of
this alternative. Finally, before concluding, in Sec. \ref{sec:4} we discuss the experimental
distinguishability of the two alternatives, and its implications for the status of becoming.

\section{The status of becoming within special relativity}\label{sec:2}

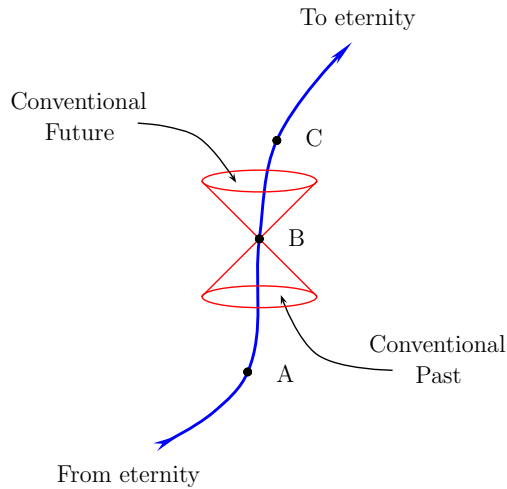
\begin{figure}
\hrule
\scalebox{0.77}{
\begin{pspicture}(0.0,-0.7)(14,8.5)

\pscurve[linewidth=0.5mm,linecolor=blue,arrowlength=3,arrowinset=0.3]{>->}(5.7,0.4)
(6.7,1.0)(7.3,1.7)(7.5,4.0)(7.8,5.7)(9.1,7.4)

\psline[linewidth=0.24mm,linecolor=red](6.5,5)(8.5,3)

\psline[linewidth=0.24mm,linecolor=red](6.5,3)(8.5,5)

\pscurve[linewidth=0.24mm,arrowsize=2pt 2]{<-}(7.87,3.03)(8.5,2.0)(9.8,1.73)

\psellipse[linewidth=0.24mm,linecolor=red](7.5,3)(1,0.2)

\psellipse[linewidth=0.24mm,linecolor=red](7.5,5)(1,0.2)

\pscurve[linewidth=0.24mm,arrowsize=2pt 2]{->}(5.4,6.0)(5.85,5.95)(6.45,5.75)(7.1,4.97)

\psdots[fillcolor=black, dotstyle=o, dotscale=1.2](7.3,1.7)

\psdots[fillcolor=black, dotstyle=o, dotscale=1.2](7.5,4.0)

\psdots[fillcolor=black, dotstyle=o, dotscale=1.2](7.8,5.7)

\put(4.0,-0.2){\large From eternity}

\put(9.4,2.05){\large Conventional}

\put(10.2,1.5){\large Past}

\put(8.2,7.7){\large To eternity}

\put(3.2,6.25){\large Conventional}

\put(3.8,5.7){\large Future}

\put(7.8,1.53){\large A}

\put(8.0,3.87){\large B}

\put(8.3,5.6){\large C}

\end{pspicture}}
\hrule
\caption{Timelike worldline of an observer tracing through an event B in a Minkowski spacetime.
Events A and C in the conventional past and conventional future of the event B are related to
B by the transitive, asymmetric, and irreflexive relation {\it precedes}. Such a linear ordering
of events is preserved under Lorentz transformations.} 

\label{figure-1}

\end{figure}

The prevalent theory of the local inertial structure at the heart of modern physics---classical
or quantal, non-gravitational or gravitational---is, of course, Einstein's special theory of
relativity. This theory, however, happens to be oblivious to any {\it structural} distinction
between the past and the future \cite{Davies-1974}. To be sure, one frequently comes across
references within its formalism to the notions of ``absolute past'' and ``absolute future''
of a given event. But these are mere conventional choices, corresponding to assignment of
tenseless linear ordering to ``static'' moments mentioned above, with the ordering now being
along the timelike worldline of an ideal observer tracing through that event
(see Fig. \ref{figure-1}).
There is, of course, no doubt about the objectivity of this ordering. It is preserved under
Lorentz transformations, and hence remains unaltered for all inertial observers. But such a
sequence of moments has little to do with becoming {\it per se}, as both physically and
mathematically well appreciated by Newton \cite{Hall and Hall-1962}\cite{Arthur-1995}, and
conceptually much clarified by McTaggart \cite{McTaggart-1908}. Worse still, there is no such
thing as a world-wide moment ``now'' in special relativity, let alone the notion of a
{\it passage} of that moment. Due to the relativity of simultaneity, what is a ``now-slice''
cutting through a given event for one observer would be a ``then-slice'' for another one moving
relative to the first, and vise versa. In other words, what is {\it past} (or has ``already
happened'') for one observer could be the {\it future} (or has ``not yet happened'') for the
other, and vise versa \cite{Penrose-1989}. This indeterminacy in temporal order cannot lead to
any causal inconsistency however, for it can only occur for spacelike separated events---i.e.,
for pairs of events lying outside the light-cones of each other. Nevertheless, these facts
suggest two rival interpretations for the continuum of events presupposed by special relativity:
(1) an {\it absolute being} interpretation and (2) a {\it relative becoming} interpretation.
According to the first of these interpretations, events in the past, present, and future exist
all at once, {\it with equal ontological status}, across the whole span of time; whereas
according to the second, events can be {\it partitioned}, causally, consistently, and
ontologically, into the sets of definite past and indefinite future events, mediated by a
fleeting present, albeit only in a relative and observer-dependent manner.

The first of these two interpretations of special relativity is sometimes also referred to as
the ``block universe'' interpretation, because of its resemblance to a 4-dimensional block of
``already laid out'' events. The moments of time in this block are supposed to be no less actual
than the locations in space are. Just as London and New York are supposed to be {\it there} even
if you may not be {\it at} either of these locations, the moments of your birth and death are
``there'' on your time-line, even if you are  presently far from being ``at'' either of these
two moments of your life. More precisely, 
along your timelike worldline all events of your life are fixed once and for all, beyond your
control, and in apparent conflict with your freedom of choice. In fact, in special relativity,
a congruence of such non-intersecting timelike worldlines---sometimes referred to as a
fibration of spacetime---represents a 3-dimensional relative space (or an inertial frame).
The 4-dimensional spacetime is then simply filled by these ``lifeless'' fibers, with the proper
time along any one of them representing the local time associated with the ordered series of
events laid out along that fiber. Informally, such a fiber is a track in spacetime of an
observer moving subluminally {\it for all eternity}. In particular, for a given moment,
all the future instants of time along this track---in exactly the same sense as all the
past instants---are supposed to be fixed, once and for all, till eternity.

Such an interpretation of time in special relativity, of course, sharply differs from our
everyday conception of time, where we expect the nonexistent future instants to {\it spring
into existence} from nowhere, streaming-in one after another, and then {\it slipping away} into
the unalterable past, thus gradually materializing the past track of our worldline, as depicted
in Fig. \ref{figure-2}. In other words, in our everyday life we normally do not think of the
future segment of our worldline to be preexisting for all eternity; instead, we perceive the
events in our lives to be occurring {\it non-fatalistically}, one after another, rendering our
worldline to ``grow'', like a tendril on a wall. But such a ``dynamic'' conception of time
appears to be completely alien to the universe purported by special relativity. Within the
Minkowski universe, as Einstein himself has been quoted as saying, ``the becoming in
three-dimensional space is transformed into a being in the world of four dimensions''
\cite{Meyerson-1985}. More famously, Weyl has gone one step further in endorsing such a static
view of the world: ``The objective world simply {\it is}, it does not {\it happen}''
\cite{Weyl-1949}. Accordingly, the appearances of change and becoming are construed to be mere
figments of our conscious experience, as Weyl
goes on to explain: ``Only to the gaze of my consciousness, crawling upward along the life line
of my body, does a section of this world come to life as a fleeting image in space which
continuously changes in time.'' Not surprisingly, some commentators have reacted strongly
against such a grim view of reality:

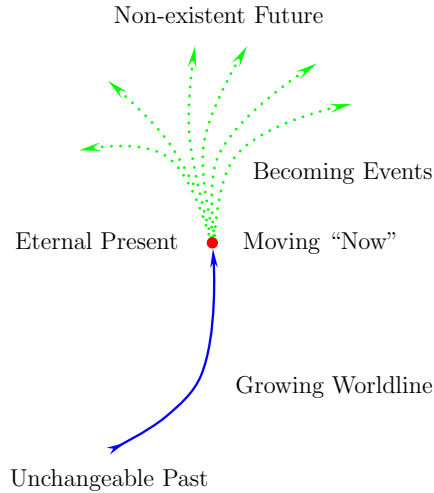
\begin{figure}
\hrule
\scalebox{0.77}{
\begin{pspicture}(0.0,-0.7)(14,8.5)

\pscurve[linewidth=0.4mm,linecolor=blue,arrowlength=2.5,arrowinset=0.3]{>->}(5.7,0.4)(6.7,1.0)(7.3,1.7)(7.51,3.9)

\pscurve[linewidth=0.5mm, linestyle=dotted, linecolor=green,arrowlength=2.5]{->}(7.53,3.84)(7.9,5.5)(9.9,6.4)
\pscurve[linewidth=0.5mm, linestyle=dotted, linecolor=green,arrowlength=2.5]{->}(7.535,4.08)(7.6,5.5)(9.3,7.1)
\pscurve[linewidth=0.5mm, linestyle=dotted, linecolor=green,arrowlength=2.5]{->}(7.55,4.0)(7.35,5.7)(8.1,7.4)
\pscurve[linewidth=0.5mm, linestyle=dotted, linecolor=green,arrowlength=2.5]{->}(7.515,3.97)(7.1,5.9)(7.2,7.4)
\pscurve[linewidth=0.5mm, linestyle=dotted, linecolor=green,arrowlength=2.5]{->}(7.23,4.9)(6.7,5.8)(5.7,6.8)
\pscurve[linewidth=0.5mm, linestyle=dotted, linecolor=green,arrowlength=2.5]{->}(7.515,3.87)(6.7,5.5)(5.2,5.6)

\psdots*[linecolor=red, fillcolor=red, dotstyle=o, dotscale=1.5](7.5,4.0)

\put(4.0,-0.2){\large Unchangeable Past}

\put(5.8,7.8){\large Non-existent Future}

\put(8.2,5.1){\large Becoming Events}

\put(8.03,3.9){\large Moving ``Now''}

\put(4.1,3.9){\large Eternal Present}

\put(7.9,1.4){\large Growing Worldline}

\end{pspicture}}
\hrule
\caption{The tensed time of the proverbial man in the street, with a degree in special
relativity. His sensation of time is much richer than a mere tenseless linear ordering of
events. Future events beyond the moving present are non-existent to him, whereas he, at
least, has a memory of the past events that have occurred along his worldline.}

\label{figure-2}
\end{figure}

\begin{Quote}
But this picture of
a ``block universe'', composed of a timeless web of ``world-lines'' in a four-dimensional space,
however strongly suggested by the theory of relativity, is a piece of gratuitous metaphysics.
Since the concept of change, of something happening, is an inseparable component of the
common-sense concept of time and a necessary component of the scientist's view of reality, it is
quite out of the question that theoretical physics should require us to hold the Eleatic view
that nothing happens in ``the objective world.'' Here, as so often in the philosophy of science,
a useful limitation in the form  of representation is mistaken for a deficiency of the
universe \cite{Black-1962}.
\end{Quote}

The frustration behind these sentiments is, of course, quite understandable. It turns out
not to be impossible, however, to appease the sentiments to some extent. It turns out that
a formal ``becoming relation'' of a limited kind can indeed be defined along a timelike
worldline, {\it uniquely} and {\it invariantly}, without in any way compromising the
principles of special relativity. The essential idea of such a relation goes back to Putnam
\cite{Putnam-1967}, who tried to demonstrate that no meaningful binary relation between two
events can exist within the framework of special relativity that can ontologically partition
a worldline into distinct parts of {\it already settled past} and {\it not yet settled future}.
Provoked by this and related arguments by Rietdijk \cite{Rietdijk-1966} and Maxwell
\cite{Maxwell-1985}, Stein \cite{Stein-1968}\cite{Stein-1991} set out to expose the
inconsistencies
within such arguments (without unduly leaning on either side of the debate), and proved that a
transitive, reflexive, and asymmetric ``becoming relation'' of a formal nature can indeed be
defined consistently between causally connected pairs of events, on a time-orientable Minkowski
spacetime. Stein's analysis has been endorsed by Shimony \cite{Shimony-now} in an approach
that is different in emphasis but complementary in philosophy, and extended by Clifton and
Hogarth \cite{Clifton-1995} to a more natural setting for the becoming along timelike worldlines.
This coherent set of arguments, taken individually or collectively, amounts to formally
proving the permissibility of objective becoming within the framework of special relativity,
but only {\it relative} to a given timelike worldline. And since a
timelike worldline in Minkowski spacetime is simply the integral curve of a never vanishing,
future-directed, timelike vector field representing the direction of a moving observer, the
becoming defended here is meaningful only {\it relative} to such an observer. There is, of
course, no inconsistency in this relativization of becoming, since---thanks to the
{\it absoluteness} of simultaneity for coincident events---different observers would always
agree on which events have already ``become'', and which have not, when their worldlines happen
to intersect. Consequently, this body of works make it abundantly plain that special relativity
{\it does not} compel us to adopt an interpretation as radical as the block universe
interpretation, but leaves room for a rather sophisticated version of our common-sense
conception of becoming. To be sure, this counterintuitive notion of a worldline-{\it dependent}
becoming permitted within special relativity is a far cry from our everyday experience, where a
{\it world-wide} present seems to perpetually stream-in from a non-existent future, and then slip
away into the unchanging past. But such a pre-relativistic notion of absolute, world-wide
becoming, occurring simultaneously for each and every one of us regardless of our motion, has no
place in the post-relativistic physics. Moreover, this apparent absolute becoming can be easily
accounted for as a gross collective of ``local'' or ``individual'' becomings along timelike
worldlines, emerging cohesively in the nonrelativistic limit. Just as Newtonian mechanics can be
viewed as an excellent approximation to the relativistic mechanics for small velocities, our
commonly shared ``world-wide'' becoming can be shown to be an excellent approximation to these
relativistic becomings for small distances, thanks to the enormity of the speed of light in
everyday units. Consequently, the true choice within special relativity should be taken not as
between absolute being and absolute becoming, but between the former (i.e., block universe) and
the {\it relativity of distant becoming}.

There has been rather surprising reluctance to accept this relativization of becoming, largely
by the proponents of the block universe interpretation of special relativity. As brought out by
Stein \cite{Stein-1991}, some of this reluctance stems from elementary misconceptions regarding
the true physical import of the theory, even by philosophers with considerable scientific
prowess. There seems to remain a genuine concern, however, because the notion of
worldline-dependent becoming tends to go against our pre-relativistic ideas of existence.
This concern can be traced back to G\"odel, who flatly refused to accept the
relativity of distant becoming on such grounds: ``A relative lapse of time, ... if any meaning
at all can be given to this phrase, would certainly be something entirely different from the
lapse of time in the ordinary sense, which means a change in the existing. The concept of
existence, however, cannot be relativized without destroying its meaning completely''
\cite{Godel-1949}. In the similar vein, in a certain book-review Callender remarks: ``... the
relativity of simultaneity poses a problem: existence itself must be relativized to frame. This
{\it may} not be a contradiction, but it is certainly a queer position to hold''
\cite{Callender-1997}. Perhaps. But nature cannot be held hostage to what our pre-relativistic
prejudices find queer. Whether we like it or not, the Newtonian notion of absolute world-wide
existence has no causal meaning in the post-relativistic physics. Within special relativity,
discernibility of events existing at a distance is constrained by the absolute upper-bound on
the speeds of causal propagation, and hence the Newtonian notion of absolute distant existence
becomes causally meaningless. To be sure, when we regress back to our everyday Euclidean
intuitions concerning the causal structure of the world, the idea of relativized existence seems
strange. However, according to special relativity the topology of this causal structure---i.e.,
the neighborhood relations between causally admissible events---happens {\it not} to be
Euclidean but {\it pseudo}-Euclidean. Once this aspect of the theory is accepted, it is quite
anomalous to hang on to the Euclidean notion of existence, or equivalently to the absoluteness
of distant becoming. It is of course logically possible to accept the relativity of distant
simultaneity but reject the relativity of distant becoming, as G\"odel seems to have done,
but conceptually that would be quite inconsistent, since the former relativity appears to
us no less queer than the latter. In fact, perhaps unwittingly, some textbook descriptions
of the relativity of simultaneity explicitly end up using the language of becoming. Witness for
example Feynman's description of a typical scenario \cite{Feynman-1963}: `` ... events that
{\it occur} at two separated places at the same time, as seen by Moe in ${S'}$, do not
{\it happen} at the same time as viewed by Joe in ${S}$ [emphasis rearranged].'' Indeed,
keeping the geometrical formalism intact, every statement involving the relativity of distant
simultaneity in special relativity can be replaced by an identical statement involving
the relativity of distant {\it becoming}, without affecting either the theoretical or the
empirical content of the theory. In other words, Einstein could have written his theory using
the latter relativity rather than the former, and that would have made no difference to the
relativistic physics---classical or quantal---of the past hundred years. The former would
have been then seen as a useful but trivial corollary of the latter. Thus, as Callender so
rightly suspects, there is indeed no contradiction in taking the relativity of distant becoming
seriously, since any evidence of our perceived co-becoming of objectively existing distant
events (i.e., of our perceived {\it absoluteness} of becoming) is quite indirect and causal
\cite{Shimony-now}. Therefore, the alleged queerness of the relativity of distant becoming
by itself cannot be taken as a good reason to opt for an interpretation of special relativity
as outrageous as the block universe interpretation.

There do exist other good reasons, however, that, on balance, land the block interpretation
the popularity it enjoys. Einstein-Minkowski spacetime is pretty ``lifeless'' on its own, as
evident from comparisons of Figs. \ref{figure-1} and \ref{figure-2} above. If becoming is a
truly ontological feature of the world, however, then we expect the sum total of reality to grow
incessantly, by objective accretion of entirely newborn events. We expect this to happen
as non-existent future events momentarily come to be the present event, and then slip away
into the unchanging past, as we saw in Fig. \ref{figure-2}. No such objective growth of reality
can be found within the Einstein-Minkowski framework for the causal structure. It is all very
well for Stein to prove the definability of a two-place ``becoming relation'' within Minkowski
spacetime, but in a genuinely becoming universe no such relation between future events and a
present event can be meaningful. Indeed, as recognize by Broad long ago, ``...the essence of a
present event is, not that it precedes future events, but that there is quite literally
{\it nothing}
to which it has the relation of precedence'' \cite{Broad-1923}. Even more tellingly, in the
Einstein-Minkowski framework there is no {\it causal compulsion} for becoming. In a genuinely
becoming universe we would expect the accretion of new events to be necessitated {\it causally},
not left at the mercy of our interpretive preferences. In other words, we would expect the entire
spatio-temporal structure to not only grow, but this growth to be also necessitated by causality
itself. No such causal dictate to becoming is there in the Einstein-Minkowski framework of
causality. A theory of local inertial structure with just such a causal necessity for objective
temporal becoming is the subject matter of our next section.

\section{A purely Heraclitean generalization of relativity}\label{sec:3}

Despite the fact that temporal transience is one of the most immediate and constantly encountered
aspects of the world \cite{Schlesinger}, Newton appears to be the last person to have actively
sought to capture it, at the most fundamental level, within a successful physical theory.
Equipped with his hypothetico-deductive methodology, he was not afraid to introduce metaphysical
notions into his theories as long as they gave rise to testable experimental consequences. After
the advent of excessively operationalistic trends within physics since the dawn of the last
century, however, questions of metaphysical flavor---questions even as important as those
concerning time---have tended to remain on the fringe of serious physical
considerations.\footnote{There are,
of course, a few brave-hearts, such as Shimony \cite{Shimony-Penrose} and Elitzur \cite{Elitzur},
who have time and again urged the physics community to take temporal becoming seriously. However,
there are also those who have preferred to explain it away as a counterfeit, resulting from
some sort of ``macroscopic irreversibility'' \cite{Primas}\cite{Hartle}\cite{Ellis}.}\label{ftn1}
Perhaps this explains why most of the popular approaches to the supposed quantum gravity are
entirely oblivious to the profound controversies concerning the status of temporal
becoming.\footnote{A welcome exception is the causal set approach initiated by Sorkin
\cite{Sorkin-2000}. However, the stochasticity of ``growth dynamics'' discovered
{\it a posteriori} within this discrete approach is a far cry from the inevitable continuity
of becoming recognized by Newton \cite{Arthur-1995}. Such a deficiency seems unavoidable
within any discrete approach to quantum gravity, due to the ``inverse problem'' of recovering
the continuum \cite{inverse}.}\label{ftn2} If, however, temporal becoming is
indeed a genuinely ontological attribute of the world, then no approach to quantum gravity can
afford to ignore it. After all, by quantum gravity one usually means a {\it complete} theory of
nature. How can a complete theory of nature be oblivious to one of the most immediate and
ubiquitous features of the world? Worse still: if temporal becoming is a genuine feature of the
world, then how can any approach to quantum gravity possibly hope to succeed while remaining in
total denial of its reality?

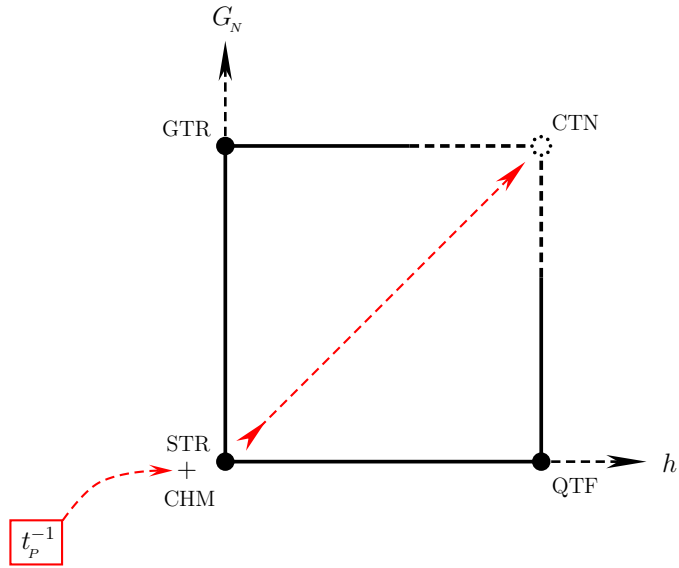
\begin{figure}
\hrule
\scalebox{0.7}{
\begin{pspicture}(-3.85,-2.5)(10,9.3)

\psline[linewidth=0.5mm,linestyle=dashed,arrowinset=0.3,arrowsize=3pt 3,arrowlength=3]{->}(2,0)
(10,0)

\psline[linewidth=0.5mm,linestyle=dashed,arrowinset=0.3,arrowsize=3pt 3,arrowlength=3]{->}(2,0)
(2,8)

\put(1.75,8.3){\Large ${\cs GN}$}

\put(10.3,-0.2){\Large ${h}$}

\put(0.8,6.2){\large ${\rm GTR}$}

\put(0.88,0.2){\large ${\rm STR}$}

\put(0.8,-0.3){\Large ${\;\;{\rm +}}$}

\put(0.83,-0.84){\large ${{\rm CHM}}$}

\put(8.2,6.3){\large ${\rm CTN}$}

\put(8.2,-0.6){\large ${\rm QTF}$}

\put(-2.1,-1.65){\Large \psframebox[linewidth=0.4mm,linecolor=red]{${\,\css t{-1}P}$}}

\pscurve[linewidth=0.4mm,linecolor=red,linestyle=dashed,arrowsize=3pt 3,arrowlength=2]{->}(-1.09,-1.12)(-0.4,-0.4)(1.0,-0.17)

\psline[linewidth=0.4mm,linecolor=red,linestyle=dashed,arrowinset=0.3,arrowsize=3pt 4,arrowlength=2.5]{>->}(2.3,0.3)(7.7,5.7)

\psline[linewidth=0.7mm,dotsize=2pt 4]{*-*}(2,0)(8,0)

\psline[linewidth=0.7mm](2,0)(2,6)

\psline[linewidth=0.7mm,dotsize=2pt 4]{*-}(2,6)(5.5,6)

\psline[linewidth=0.7mm,linestyle=dashed]{-}(5.5,6)(7.7,6)

\pscircle[linewidth=0.7mm,linestyle=dotted](8,6){6pt}

\psline[linewidth=0.7mm](8,0)(8,3.5)

\psline[linewidth=0.7mm,linestyle=dashed](8,3.5)(8,5.72)

\end{pspicture}}
\hrule
\caption{Introducing the inverse of the Planck time at the conjunction of Special Theory
of Relativity (STR) and Classical Hamiltonian Mechanics (CHM), with a bottom-up view to
a Complete Theory of Nature (CTN). Both General Theory of Relativity (GTR) and Quantum
Theory of Fields (QTF) are viewed as limiting cases, corresponding to negligible quantum
effects (represented by Planck's constant ${h}$) and negligible gravitational effects
(represented by Newton's constant ${\cs GN}$), respectively.}

\label{figure-3}

\end{figure}

Partly in response to such ontological and methodological questions, an intrinsically
Heraclitean generalization of special relativity was constructed in Ref. \cite{Christian-2004}.
The strategy behind this approach was to judiciously introduce the inverse of the Planck time,
namely ${\css t{-1}P}$, at the {\it conjunction} of special relativity and Hamiltonian
mechanics, with a bottom-up view to a {\it complete theory of nature}, in a manner similar to
how general relativity was erected by Einstein on special relativity (see Fig. \ref{figure-3}).
The resulting theory of the causal
structure has already exhibited some remarkable physical consequences. In particular, such a
judicious introduction of ${\css t{-1}P}$ necessitates energies and momenta to be invariantly
bounded from above, and lengths and durations similarly bounded from below, by their respective
Planck scale values. By contrast, within special relativity nothing prevents physical quantities
such as energies and momenta to become unphysically large---i.e., infinite---in a rapidly
moving frame. In view of the primary purpose of the present essay, however, we shall refrain
form dwelling too much into these physical consequences (details of which may be found in Ref.
\cite{Christian-2004}). Instead, we shall focus here on those features of the generalized
theory that accentuate its purely Heraclitean character.

\begin{figure}
\hrule
\scalebox{0.75}{
\begin{pspicture}(-0.2,8)(14,15.4)

\pscurve[linewidth=0.5mm]{**-**}(2.1,10.3)(2.8,11.1)(4.7,11.5)(5.6,13)

\pscurve[linewidth=0.5mm](1.45,9.55)(6.2,9.32)(6.3,11.5)(6.5,14.5)
(4,13.7)(1.4,13.5)(1,9.9)(1.45,9.55)

\put(1.8,13){\large ${\cal M}$}

\put(5.2,9.95){\large ${4}$}

\put(1.65,9.95){\large ${{\rm e}_1}$}

\put(5.7,13.2){\large ${{\rm e}_2}$}

\pscurve[linewidth=0.5mm]{**-**}(9.7,10)(13,11.5)(11.2,11.1)(13,13)

\pscurve[linewidth=0.5mm](14,10)(14,14)(12,13.7)(9,14)(9,9)(11.2,9)(13.6,9.4)(14,10)

\put(6.9,11.5){\Huge ${\times}$}

\put(9.2,9.8){\large ${{\rm s}_1}$}

\put(13.2,13.1){\large ${{\rm s}_2}$}

\put(12.9,9.8){\large ${2n}$}

\put(9,13.1){\large ${\cal N}$}

\end{pspicture}}
\hrule
\caption{(a) The motion of a clock from event ${{\rm e}_1}$ to event ${{\rm e}_2}$ in a
Minkowski Spacetime ${\cal M}$. (b) As the clock moves from ${{\rm e}_1}$ to ${{\rm e}_2}$,
it also inevitably evolves, as a result of its external motion, from state ${{\rm s}_1}$ to
state ${{\rm s}_2}$ in its own ${2n}$-dimensional phase space ${\cal N}$.}

\label{figure-4}

\end{figure}
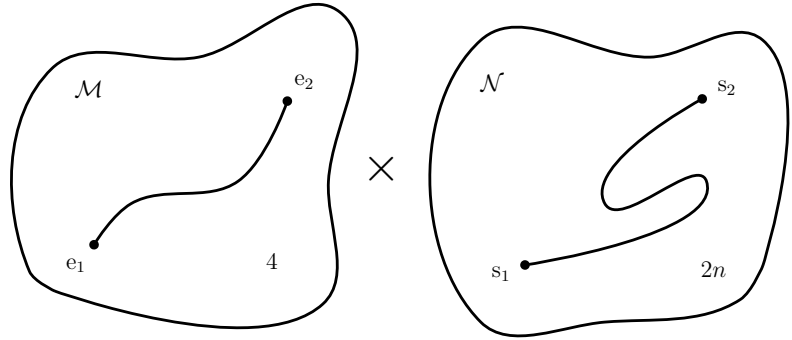

\subsection{Fresh look at the proper duration in special relativity}\label{subsec:3.1}

To this end, let us reassess the notion of {\it proper duration} residing at the very heart of
special relativity. Suppose an object system, equipped with an ideal classical clock of unlimited
accuracy, is moving with a uniform velocity ${\bf{v}}$ in a Minkowski spacetime ${\cal M}$,
from an event ${{\rm e}_1}$ at the origin of a reference frame to a nearby event ${{\rm e}_2}$
in the future light cone of ${{\rm e}_1}$, as shown in Fig. \ref{figure-4}a. For our purposes,
it would suffice to refer to this system, say of ${n}$ degrees of freedom, simply as ``the
clock.'' As it moves, the clock will also {\it necessarily} evolve, as a result of its external
motion, say at a uniform rate ${\boldsymbol{\omega}}$, from one state, say ${{\rm s}_1}$,
to another state, say ${{\rm s}_2}$, within its own relativistic phase space, say ${\cal N}$.
In other words, the inevitable evolution of the clock from ${{\rm s}_1}$ to ${{\rm s}_2}$---or
rather that of its state---will trace out a unique trajectory in the phase space ${\cal N}$,
as shown in Fig. \ref{figure-4}b. For simplicity, we shall assume that this phase space of the
clock is finite dimensional; apart from possible mathematical encumbrances, the reasoning that
follows would go through unabated for the case of infinite dimensional phase spaces (e.g. for
clocks made out of relativistic fields).

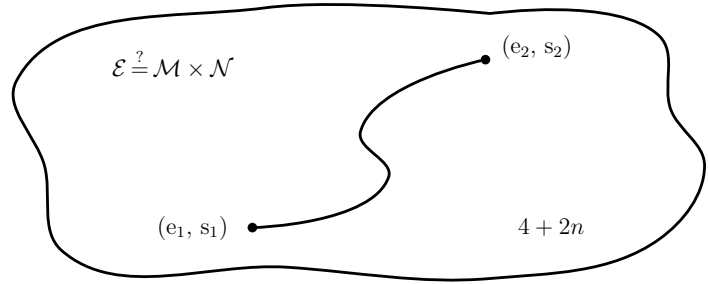
\begin{figure}
\hrule
\scalebox{0.75}{
\begin{pspicture}(-0.2,2.6)(14,9.27)

\pscurve[linewidth=0.5mm]{**-**}(5.7,4.4)(8.2,5.3)(7.7,6.1)(10,7.4)

\pscurve[linewidth=0.5mm](10,8.2)(6,8.3)(1.6,7)(2.1,5.5)(2.4,4)(6.4,3.7)
(10,3.5)(13,4.2)(13.8,5.5)(13.2,6.5)(13,7.8)(10,8.2)

\put(4.1,4.3){\large ${({\rm e}_1,\,{\rm s}_1)}$}

\put(10.2,7.5){\large ${({\rm e}_2,\,{\rm s}_2)}$}

\put(10.5,4.3){\large ${4+2n}$}

\put(3.3,7.1){\large ${{\cal E}\,{\buildrel ? \over =}\,{\cal M}\times{\cal N}}$}

\end{pspicture}}
\hrule
\caption{What is the correct metric-topology of the combined space ${\cal E}$---made up of the
external Minkowski spacetime ${\cal M}$ and the internal space of states ${\cal N}$---in
which our clock moves as well as evolves from event-state ${({\rm e}_1,\,{\rm s}_1)}$ to
event-state ${({\rm e}_2,\,{\rm s}_2)}$?}

\label{figure-5}
\end{figure}

Now, nothing prevents us from thinking of this motion and evolution of the clock conjointly,
as taking place in a combined ${4+2n-}$dimensional space, say ${\cal E}$, the elements of
which may be called {\it event-states} and represented by pairs ${({\rm e}_i,\,{\rm s}_i)}$,
as depicted in Fig. \ref{figure-5}.
Undoubtedly, it is this combined space that truly captures the complete specification
of all possible physical attributes of our classical clock. Therefore, we may ask:
{\it What will be the time interval actually registered by the clock as it
moves and evolves from the event-state ${({\rm e}_1,\,{\rm s}_1)}$ to the event-state
${({\rm e}_2,\,{\rm s}_2)}$ in this combined space ${\cal E}$}? It is
only by answering such a physical question can one determine the correct topology and
geometry of the combined space in the form of a metric, analogous to the Minkowski metric
corresponding to the line element
\begin{equation}
d{\css {\tau}2E}=dt^2 - c^{-2}d{\bf x}^2\geq 0\,,\label{Minkowski}
\end{equation}
where the inequality asserts the causality condition.
Of course, after Einstein the traditional answer to the above question, in accordance with the
line element (\ref{Minkowski}), is simply
\begin{equation}
\Delta{\cs {\tau}E}=\int_{({\rm e}_1,\,{\rm s}_1)}^{({\rm e}_2,\,{\rm s}_2)}\!d{\cs {\tau}E}=
\int_{{\rm t}_1}^{{\rm t}_2}\!\frac{1}{\gamma(v)}\;dt\,,\label{persp}
\end{equation}
with the usual Lorentz factor
\begin{equation}
\gamma(v):=\frac{1}{\sqrt{1-v^2/c^2}} > 1.\label{us-gamma}
\end{equation}
In other words, the traditional answer is that the metrical topology of the space ${\cal E}$ is
of a product form, ${{\cal E}={\cal M}\times{\cal N}}$, and---more to the point---the clock that
records the duration ${\Delta{\cs {\tau}E}}$ in question remains {\it insensitive} to the passage
of time that marks the evolution of variables within its own phase space ${\cal N}$.

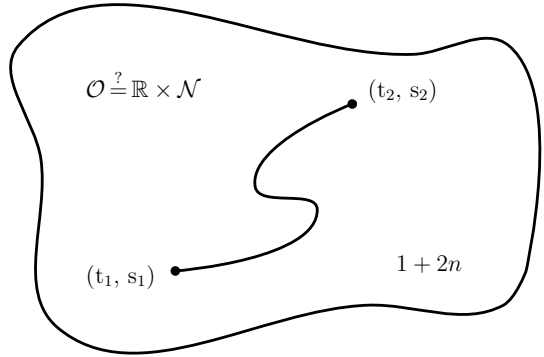
\begin{figure}
\hrule
\scalebox{0.75}{
\begin{pspicture}(3.8,7.75)(10,15.5)

\pscurve[linewidth=0.5mm]{**-**}(9.7,10)(12.3,11.1)(11.2,11.5)(13,13)

\pscurve[linewidth=0.5mm](16,10)(15.7,14)
(14,13.85)(7,14)(7.4,12)(7.7,9)(13.2,9.4)(15.6,9.4)(16,10)

\put(8.2,9.8){\large ${({\rm t}_1,\,{\rm s}_1)}$}

\put(13.2,13.1){\large ${({\rm t}_2,\,{\rm s}_2)}$}

\put(13.7,10){\large ${1+2n}$}

\put(8.2,13.1){\large ${{\cal O}\,{\buildrel ? \over =}\,\mathbb{R}\times{\cal N}}$}

\end{pspicture}}
\hrule
\caption{The evolution of our clock from instant-sate ${({\rm t}_1,\,{\rm s}_1)}$ to
instant-state ${({\rm t}_2,\,{\rm s}_2)}$ in the odd dimensional extended phase
space ${\cal O}$. What is the correct topology of ${\cal O}$?}

\label{figure-6}
\end{figure}

But from the above perspective---i.e., from the perspective of Fig. \ref{figure-5}---it is
evident that Einstein made an implicit assumption while proposing the proper duration
(\ref{persp}). He tacitly assumed that the rate at which a given physical state can evolve
remains {\it unbounded}. Of course, he had no particular reason to question the limitlessness
of how fast a physical state can evolve. However, for us---from what we have learned from our
efforts to construct a theory of quantum gravity---it is not unreasonable to suspect that the
possible rate at which a physical state can evolve is invariantly bounded from above. Indeed,
it is generally believed that the Planck scale marks a threshold beyond which our theories of
space and time, and possibly also of quantum phenomena, are unlikely to survive
\cite{Christian-2004}\cite{Christian-2001}\cite{Christian-PRL}. In particular, the Planck time
${\cs tP}$ is widely thought to be {\it the} minimum possible duration. It is then only natural
to suspect that the inverse of the Planck time---namely ${\css t{-1}P}$, with its approximate
value of ${10^{+43}}$ Hertz in ordinary units---must correspond to the absolute upper bound on
how fast a physical state can possibly evolve. In this context, it is also worth noting that the
speed of light is simply a ratio of the Planck length over the Planck time,
${c:={\cs lP}/{\cs tP}}$, which suggests that perhaps the assumption of absolute upper bound
${\css t{-1}P}$ on possible rates of evolution should be taken to be more primitive in physical
theories than the usual assumption of absolute upper bound ${c}$ on possible speeds of motion.
In fact, as we shall see, the assumption of upper bound ${c}$ on speeds of motion can indeed be
viewed as a special case of our assumption of upper bound ${\css t{-1}P}$ on rates of evolution.

To this end, let us then systematize the above thoughts by incorporating ${\css t{-1}P}$ into
a physically viable and empirically adequate theory of the local causal structure. One way
to accomplish this task is to first consider a simplified picture, represented by what is known
as the extended phase space, constructed within a global inertial frame in which the clock is
at rest (see Fig. \ref{figure-6}). Now, in such a frame the proper time interval the clock
would register is simply the Newtonian time interval ${\Delta t}$. Using this time
${t\in\mathbb{R}}$ as an external parameter, within this frame one can determine the extended
phase space ${{\cal O}=\mathbb{R}\times{\cal N}}$ for the dynamical evolution of the clock using
the usual Hamiltonian prescription. Suppose next we consider time-dependent canonical
transformations of the dimensionless phase space coordinates ${{\rm y}^{\mu}(t)}$
(${\mu=1,\ldots,2n}$), expressed in Planck units, into coordinates ${{\rm y}'^{\mu}(t)}$
of the following general linear form:
\begin{equation}
{\rm y}'^{\mu}({\rm y}^{\mu}(0),\,t)={\rm y}^{\mu}(0)+
\omega_r^{\mu}({\bf y}(0))\,t+b^{\mu}\,,\label{cano-trans}
\end{equation}
where ${\omega_r^{\mu}}$ and ${b^{\mu}}$ do not have explicit time dependence, and the
reason for the subscript ${r}$ in ${\omega_r^{\mu}}$, which stands for ``relative'', will
become clear soon. Interpreted actively, these are simply the linearized solutions of the
familiar Hamiltonian flow equations,
\begin{equation}
\frac{d{\rm y}^{\mu}}{dt\;}=\omega_r^{\mu}\left({\bf y}(t)\right)
:=\Omega^{\mu\nu}\frac{\partial H}{\partial {\rm y}^{\nu}}\,,\label{H-eqn}
\end{equation}
where ${{\boldsymbol\omega}_r}$ is the Hamiltonian vector field generating the flow,
${{\bf y}(t)}$ is a ${2n}$-dimensional local Darboux vector in the phase space
${\cal N}$, ${\boldsymbol{\Omega}}$ is the symplectic 2-form on ${\cal N}$, and ${H}$ is a
Hamiltonian function governing the evolution of the clock. If we now denote by
${\omega^{\mu}}$ the uniform time rate of change of the canonical coordinates ${{\rm y}^{\mu}}$,
then the linear transformations (\ref{cano-trans}) imply the composition law
\begin{equation}
\omega'^{\mu}=\omega_{}^{\mu}+\omega_r^{\mu}\label{omega-trans}
\end{equation}
for the evolution rates of the two sets of coordinates, with ${-\omega_r^{\mu}}$ interpreted
as the rate of evolution of the transformed coordinates with respect to the original ones.
Crucially for our purposes, what is implicit in the law (\ref{omega-trans}) is the assumption
that there is no upper bound on the rates of evolution of physical states. Indeed, successive
transformations of the type (\ref{cano-trans}) can be used, along with (\ref{omega-trans}), to
generate arbitrarily high rates of evolution for the state of the clock. More pertinently, the
assumed validity of the composition law (\ref{omega-trans}) turns out to be equivalent to
assuming the absolute simultaneity of ``instant-states'' ${({\rm t}_i,\,{\rm s}_i)}$ within
the ${1+2n-}$dimensional extended phase space ${\cal O}$. In other words, within the
${1+2n-}$dimensional manifold ${\cal O}$, the ${2n-}$dimensional phase spaces simply
constitute strata of
hypersurfaces of simultaneity, much like the strata of spatial hypersurfaces within a Newtonian
spacetime. Indeed, the extended phase spaces such as ${\cal O}$ are usually taken to be contact
manifolds, with topology presumed to be a product of the form ${\mathbb{R}\times{\cal N}}$. 

Thus, not surprisingly, the assumption of absolute time in contact spaces is equivalent to the
assumption of ``no upper bound'' on the possible rates of evolution of physical states. Now, in
accordance with our discussion above, suppose we impose the following upper bound on the
evolution rates\footnote{The ``flat'' Euclidean metric on the phase space that is being used
here is the ``quantum shadow metric'', viewed as a classical limit of the Fubini-Study metric of
the quantum state space (namely, the projective Hilbert space), in accordance with our bottom-up
philosophy depicted in Fig. \ref{figure-3}. See Ref. \cite{Christian-2004} for further
details.\label{ftn3}}:
\begin{equation}
\left|\frac{d{\bf y}}{dt}\right| =: \omega\leq {\css t{-1}P}\,.\label{t-bound}
\end{equation}
If this upper bound is to have any physical significance, however, then it must hold for
{\it all} possible evolving phase space coordinates ${{\rm y}^{\mu}(t)}$, and that is
amenable if and only if the composition law (\ref{omega-trans}) is replaced by
\begin{equation}
\omega'^{\mu}=\frac{\omega_{}^{\mu}+\,\omega^{\mu}_r}{1+{\css t2P}\;
\omega_{}^{\mu}\,\omega_r^{\mu}}\,,\label{w-relation}
\end{equation}
which implies that as long as neither ${\omega^{\mu}}$ nor ${\omega^{\mu}_r}$ exceeds the
causal upper bound ${\css t{-1}P}$, ${\omega'^{\mu}}$ also remains within ${\css t{-1}P}$.
Of course, this generalized law of composition has been inspired by Einstein's own such
law for velocities, which states that the velocity, say ${v_{}^k}$ (${k=1,2,}$ or ${3}$), of
a material body in a given direction in one inertial frame is related to its velocity, say
${v'^k}$, in another frame, moving with a velocity ${-v^k_r}$ with respect to the first, by
the relation
\begin{equation}
v'^k=\frac{v_{}^k+\,v^k_r}{1+c^{-2}\,v_{}^k\,v^k_r}\,.\label{v-relation}
\end{equation}
Thus, as long as neither ${v_{}^k}$ nor ${v^k_r}$ exceeds the upper bound
${c}$, ${v'^k}$ also remains within ${c}$. It is this absoluteness of ${c}$
that lends credence to the view that it is merely a conversion factor between
the dimensions of time and space. This fact is captured most conspicuously by
the quadratic invariant (\ref{Minkowski}) of spacetime.
In exact analogy, if we require the causal relationships among the possible instant-states
${({\rm t}_i,\,{\rm s}_i)}$ in ${\cal O}$ to respect the upper bound ${\css t{-1}P}$ in
accordance with the law (\ref{w-relation}), then the usual product metric of the space
${\cal O}$ would have to be replaced by the pseudo-Euclidean metric corresponding to the
line element
\begin{equation}
d{\css {\tau}2H}=dt^2 - {\css t2P}d{\bf{y}}^2\geq 0,\label{a-metric}
\end{equation}
where the phase space line element ${d\bf{y}}$ was discussed in the footnote \ref{ftn3} above.
But then, in the resulting picture, different canonical coordinates evolving with
nonzero relative rates would differ in general over which instant-states are simultaneous
with a given instant-state. As unorthodox as this new picture may appear to be, it is an
inevitable consequence of the upper bound (\ref{t-bound}).

Let us now raise a question analogous to the one raised earlier: In its rest frame,
what will be the time interval registered by the clock as it evolves from an instant-state
${(t_1,\,s_1)}$ to an instant-state ${(t_2,\,s_2)}$ within the space ${\cal O}$? The answer,
according to the pseudo-Euclidean line element (\ref{a-metric}), is clearly
\begin{equation}
\Delta{\cs {\tau}H}=\!\int_{({\rm t}_1,\,{\rm s}_1)}^{({\rm t}_2,\,{\rm s}_2)}\!d{\cs {\tau}H}
=\!\int_{{\rm t}_1}^{{\rm t}_2}
\!\frac{1}{\gamma(\omega)}\;dt=\frac{|\,{\rm t}_2-{\rm t}_1|}{\gamma(\omega)}
=\frac{\Delta t}{\gamma(\omega)}\,,
\end{equation}
where
\begin{equation}
\gamma(\omega):=\frac{1}{\sqrt{1-
{\css t2P}\,\omega^2}} > 1.
\end{equation}
Thus, if the state of the clock is evolving, then we will have the phenomenon of ``time
dilation'' even in the rest frame. Similarly, we will have a phenomenon of ``{\it state}
contraction'' in analogy with the phenomenon of ``length contraction'': 
\begin{equation}
\Delta {\rm y}'=\omega\,\Delta{\cs {\tau}H}=\frac{\;\omega\,\Delta t}
{\gamma(\omega)}=\frac{\Delta {\rm y}}{\gamma(\omega)}\,.
\end{equation}
It is worth emphasizing here, however, that, as in ordinary special relativity, nothing is
actually ``dilating'' or ``contracting''. All that is being exhibited by these phenomena is
that the two sets of mutually evolving canonical coordinates happen to differ over which
instant-states are simultaneous.

So far, to arrive at the expression (\ref{a-metric}) for the proper duration, we have used a
specific Lorentz frame, namely the rest frame of the clock. In a frame with respect to which
the same clock is uniformly moving, the expression for the actual proper duration can be
obtained at once from (\ref{a-metric}), by simply using the Minkowski line element
(\ref{Minkowski}), yielding
\begin{equation}
d\tau^2=dt^2 - c^{-2}d{\bf x}^2-{\css t2P}d{\bf{y}}^2\geq 0.\label{Minkowski-plus}
\end{equation}
This, then, is the ${4+2n-}$dimensional quadratic invariant of our combined space ${\cal E}$ of
Fig. \ref{figure-5}. We may now return to our original question and ask:
What, according to this generalized theory of relativity, will be the proper duration
registered by a given clock as it moves and evolves from an event-state
${({\rm e}_1,\,{\rm s}_1)}$ to an event-state ${({\rm e}_2,\,{\rm s}_2)}$ in the combined space
${\cal E}$? Evidently, according to the quadratic invariant (\ref{Minkowski-plus}), the
answer is simply:
\begin{equation}
\Delta\tau=\!\int_{({\rm e}_1,\,{\rm s}_1)}^{({\rm e}_2,\,{\rm s}_2)}\!d\tau=\!
\int_{{\rm t}_1}^{{\rm t}_2}\!\!\frac{1}{\gamma(v,\,\omega)}\,dt\,,
\end{equation}
with
\begin{equation}
\gamma(v,\,\omega):=\frac{1}{\sqrt{1-c^{-2}\,v^2 -
{\css t2P}\,\omega^2}} > 1.\label{gengamm}
\end{equation}

We are now in a position to isolate the two basic postulates on which the generalized theory
of relativity developed above can be erected in the manner analogous to the usual special
relativity. In fact, the first of the two postulates can be taken to be Einstein's very own
first postulate, except that we must now revise the meaning of inertial coordinate system. In
the present theory it is taken to be a system of ${4+2n}$ dimensions, ``moving'' uniformly in
the combined space ${\cal E}$, with 4 being the external spacetime dimensions, and ${2n}$ being
the internal phase space dimensions of the system. Again, the internal dimensions of the object
system can be either finite or infinite in number. Next, note that by eliminating the speed of
light in favor of pure Planck scale quantities the quadratic invariant (\ref{Minkowski-plus})
can be expressed in the form
\begin{equation}
d\tau^2=\,dt^2-{\css t2P}\left\{{\css l{-2}P}\,d{\bf{x}}^2+\,d{\bf{y}}^2\right\}\geq 0\,,
\label{newquad}
\end{equation}
where ${\cs lP}$ is the Planck length of the value ${\sim 10^{-33}}$ cm in ordinary units.
The two postulates of generalized relativity may now be stated as follows:
\begin{description}[label]
\item[(1)] {\it The laws governing the states of a physical system are insensitive to
``the state of motion'' of the ${4+2n-}$dimensional reference coordinate system in the
pseudo-Euclidean space ${\cal E}$, as long as it remains ``inertial''}.
\item[(2)] {\it No time rate of change of a dimensionless physical quantity, expressed in Planck
units, can exceed the inverse of the Planck time}.
\end{description}

Clearly, the generalized invariance embedded within this new causal theory of local inertial
structure is much broader in its scope---both physically and conceptually---than the
invariance embedded within special relativity. For example, in the present theory even the
four dimensional continuum of spacetime no longer enjoys the absolute status it does in
Einstein's theories of relativity. Einstein dislodged Newtonian concepts of absolute time and
absolute space, only to replace them by an analogous concept of {\it absolute
spacetime}---namely, a continuum of {\it in principle} observable events, idealized as a
connected pseudo-Riemannian manifold, with observer-independent spacetime intervals. Since it is
impossible to directly observe this remaining absolute structure without recourse to the
behavior of material objects, perhaps it is best viewed as the ``ether'' of the modern times, as
Einstein himself occasionally did \cite{Einstein-1920}. By contrast, it is evident that in the
present theory even this four-dimensional spacetime continuum has no absolute,
observer-independent meaning. In fact, apart from the laws of nature, there is very little
absolute structure left in the present theory, for now even the quadratic invariant
(\ref{Minkowski-plus}) is dependent on the phase space structure of the material system
being employed.

\subsection{Physical implications of the generalized theory of relativity}\label{subsec:3.2}

Although not our main concern here\footnote{In this subsection we shall only briefly highlight
the physical implications of the generalized theory of relativity. For a complete discussion
see Sec. VI of Ref. \cite{Christian-2004}.}, it is worth noting that the generalized theory
of relativity described above is both a physically viable and empirically adequate theory. In
fact, in several respects the present theory happens to be physically better behaved than
Einstein's special theory of relativity. For instance, unlike in special relativity, in the
present theory physical quantities such as lengths, durations, energies, and momenta remain
bounded by their respective Planck scale values. This physically sensible behavior is due to
the fact that present theory assumes even less preferred structure than special relativity,
by positing democracy among the internal phase space coordinates in addition to that
among the external spacetime coordinates.

Mathematically, this demand of combined democracy among spacetime and phase space
coordinates can be captured by requiring invariance of the physical laws under
the ${4+2n-}$dimensional coordinate transformations \cite{Christian-2004}
\begin{equation}
z^A=\Lambda^A_{\,\;\;B}\,z'^B +\,b^A\label{transla}
\end{equation}
analogous to the Poincar\'e transformations, with the index
${A=0,\dots,3+2n}$ now running along the ${4+2n}$ dimensions of the
manifold ${\cal E}$ of Fig. \ref{figure-5}. These transformations would preserve the
quadratic invariant (\ref{newquad}) iff the constraints
\begin{equation}
\Lambda^A_{\,\;\;C}\,\Lambda^B_{\,\;\;D}\;\xi_{AB}=\xi_{CD}\label{consla}
\end{equation}
are satisfied, where ${\xi_{AB}}$ are the components of the metric on the manifold
${\cal E}$. At least for simple finite dimensional phase spaces, the coefficients
${\Lambda^A_{\,\;\;B}}$ are easily determinable. For example, consider a massive relativistic
particle at rest (and hence also not evolving) with respect to a primed coordinate system in
the external spacetime, which is moving with a uniform velocity ${\bf v}$ with respect to
another unprimed coordinate system. Since, as it moves, the state of the particle will also
be evolving in its six dimensional phase space, say at a uniform rate ${\boldsymbol{\omega}}$,
we can view its motion and evolution together with respect to a ${4+6-}$dimensional unprimed
coordinate system in the space ${\cal E}$.

Restricting now to the external spatio-temporal sector where we actually perform our
measurements, it is easy to show
\cite{Christian-2004} that the coefficients ${\Lambda^A_{\,\;\;B}}$ are functions of the
generalized gamma factor (\ref{gengamm}), with the corresponding expression for the length
contraction being
\begin{equation}
\Delta {\rm x}'=\frac{\Delta {\rm x}}{\gamma(v,\,\omega)}\,,\label{lc}
\end{equation}
which can be further evaluated to yield
\begin{equation}
\Delta {\rm x}'=\Delta {\rm x}\,\sqrt{1-\frac{v^2}{c^2} -
\,{\css l2P}\left(\frac{{\Delta {\rm x}}-{\Delta {\rm x}'}}{{\Delta {\rm x}'}{\Delta {\rm x}}}
\right)^2}\,.\label{dist}
\end{equation}
Although nonlinear, this expression evidently reduces to the special relativistic expression for
length contraction in the limit of vanishing Planck length. For the physically interesting case
of ${\Delta {\rm x}'\ll\Delta {\rm x}}$, it can be simplified and solved exactly, yielding the
``linearized'' expression for the ``contracted'' length,
\begin{equation}
\Delta {\rm x}'=\Delta {\rm x}\,\sqrt{\frac{1}{2}\left(1-\frac{v^2}{c^2}\right) +
{\sqrt{\frac{1}{4}\left(1-\frac{v^2}{c^2}\right)^{\!2} -
\frac{{\css l2P}}{(\Delta {\rm x})^2}}}}\;,\label{l}
\end{equation}
provided the reality condition
\begin{equation}
\frac{1}{4}\left(1-\frac{v^2}{c^2}\right)^{\!2}\;\geq\;
\frac{{\css l2P}}{(\Delta {\rm x})^2}\label{inequal}
\end{equation}
is satisfied. Substituting this condition back into the solution (\ref{l}) then gives
\begin{equation}
\Delta {\rm x}'\;\geq\;\sqrt{{\cs lP}\Delta {\rm x}}\;,\label{bound-pre}
\end{equation}
which implies that as long as ${\Delta {\rm x}}$ remains greater than ${\cs lP}$ the
``contracted'' length ${\Delta {\rm x}'}$ also remains greater than ${\cs lP}$, in close
analogy with the invariant bound ${c}$ on speeds in special relativity. That is to say, in
addition to the upper bound ${\Delta {\rm x}}$ on lengths implied by the condition
${\gamma(v,\,\omega)>1}$ above, the ``contracted'' length ${\Delta {\rm x}'}$ also remains
invariantly bounded from below, by ${\cs lP}$:
\begin{equation}
\Delta {\rm x}\;>\;\Delta {\rm x}'\;>\;{\cs lP}\,.\label{bound-l}
\end{equation}

Starting again from the expression for time dilation analogous to that for the length
contraction,
\begin{equation}
\Delta\tau=\frac{\Delta t}{\gamma(v,\,\omega)}\,,\label{ltc}
\end{equation}
and using almost identical line of arguments as above, one analogously arrives at a generalized
expression for the time dilation,
\begin{equation}
\Delta\tau=\Delta t\,\sqrt{\frac{1}{2}\left(1-\frac{v^2}{c^2}\right) +
{\sqrt{\frac{1}{4}\left(1-\frac{v^2}{c^2}\right)^{\!2} -
\frac{{\css t2P}}{(\Delta t)^2}}}}\;,\label{t}
\end{equation}
together with the corresponding invariant bounds on the ``dilated'' time:
\begin{equation}
\Delta t\;>\;\Delta\tau\;>\;{\cs tP}\,.\label{bound-t}
\end{equation}
Thus, in addition to being bounded from above by the time ${\Delta t}$, the ``dilated'' time
${\Delta \tau}$ remains invariantly bounded also from below, by the Planck time ${\cs tP}$. 

So far we have not assumed or proved explicitly that the constant ``${c}$'' is an upper bound
on possible speeds. As alluded to above, in the present theory the observer-independence of the
upper bound ${c}$ turns out to be a derivative notion. This can be easily appreciated by
considering the ratio of the ``contracted'' length (\ref{l}) and ``dilated'' time (\ref{t}),
along with the definitions
\begin{equation}
u:=\frac{\Delta {\rm x}}{\Delta t}\;\;\;\;\;\;{\rm and}\;\;\;\;\;\;
u':=\frac{\Delta {\rm x}'}{\Delta\tau}\label{u-defined}
\end{equation}
for velocities, leading to the upper bound on velocities in the moving frame:
\begin{equation}
u'\,\leq\,u\,\sqrt{1+\sqrt{1-c^2\,u^{-2}}}\,.\label{reaffirmed}
\end{equation}
Hence, as long as ${u}$ does not exceed ${c}$, ${u'}$ also remains within ${c}$. In other words,
in the present theory ${c}$ retains its usual status of the observer-independent upper bound on
causally admissible speeds, but in a rather derivative manner.

In addition to the above kinematical implications, the basic elements of the particle physics
are also modified within our generalized theory, the central among which being
the Planck scale ameliorated dispersion relation
\begin{equation}
p^2\,c^2\,+\;m^2\,c^4\,=\,E^2\left[\,1\,-\,\frac{\;\left(E-m\,c^2\right)^2}{\css E2P}\,\right],
\label{exact-disp}
\end{equation}
where ${\cs EP}$ is the Planck energy. It is worth emphasizing here that this is an {\it exact}
relation between energies and momenta, which in the rest frame of the massive particle
reproduces Einstein's famous mass-energy equivalence:
\begin{equation}
E=m\,c^2.\label{mass-energy}
\end{equation}
Moreover, in analogy with the invariant lower bounds on lengths and durations we discussed above,
in the present theory energies and momenta can also be shown to remain invariantly bounded from
above by their Planck values:
\begin{equation}
E'\;\leq\;\sqrt{{\cs EP}E}\;\;\;\;\;{\rm and}\;\;\;\;\;
p'\;\leq\;\sqrt{{\cs kP}\,p}\,\times\sqrt{\frac{v}{c}}\;,
\label{E-and-p-bound}
\end{equation}
where ${\cs kP}$ is the Planck momentum.
Thus, as long as the unprimed energy ${E}$ does not exceed ${\cs EP}$, the primed energy ${E'}$
also remains within ${\cs EP}$. That is to say, in addition to the lower bound ${E}$ on energies
implied by the condition ${\gamma(v,\,\omega)>1}$, the energies ${E'}$ remain invariantly
bounded also from above, by the Planck energy ${\cs EP}$:
\begin{equation}
E\;<\;E'\;<\;{\cs EP}\,.\label{bound-e}
\end{equation}
Similarly, as long as the relative velocity ${v}$ does not exceed ${c}$ and the unprimed momentum
${p}$ does not exceed ${\cs kP}$, the primed momentum ${p'}$ also remains within ${\cs kP}$.
Hence, in addition to the lower bound ${p}$ on momenta set by the condition
${\gamma(v,\,\omega)>1}$, the momenta ${p'}$ remain invariantly bounded also from above, by
the Planck momentum ${\cs kP}$:
\begin{equation}
p\;<\;p'\;<\;{\cs kP}\,.\label{bound-p}
\end{equation}
Thus, unlike in special relativity, in the present theory {\it all} physical quantities remain
invariantly bounded by their respective Planck scale values.

Next, consider an isolated system of mass ${m_{sys}}$ composed of a number of constituents
undergoing an internal reaction. It follows from the quadratic invariant (\ref{newquad}) that
the ${4+2n-}$vector ${{\boldsymbol{\cal P}}_{\!sys}}$, defined as the abstract momentum of the
system as a whole, would be conserved in such a reaction (cf. \cite{Christian-2004}),
\begin{equation}
\Delta {\boldsymbol{\cal P}}_{\!sys}=0\,,\label{totmomcons}
\end{equation}
where ${\Delta}$ denotes the difference between initial and final states of the reaction,
and ${{\boldsymbol{\cal P}}_{\!sys}}$ is defined by
\begin{equation}
m_{sys}\,\frac{\;dz^A}{d\tau\,}=:{\cal P}_{\!sys}^A := 
\left(E_{sys}/c\,,\;p_{sys}^k\,,\;P^{\mu}_{sys}\right),\label{4+2n-sys}
\end{equation}
with ${k=1,2,3}$ denoting the external three dimensions and ${\mu=4,5,\dots,3+2n}$ denoting the
phase space dimensions of the system as a whole. It is clear from this definition that, since
${dz^A}$ is a ${4+2n-}$vector whereas ${m_{sys}}$ and ${d\tau}$ are invariants,
${{\cal P}_{\!sys}^A}$ is also a ${4+2n-}$vector, and hence transforms under (\ref{transla}) as
\begin{equation}
{\cal P}_{\!sys}'^A=\Lambda^A_{\,\;\;B}\;{\cal P}_{\!sys}^B\,.\label{transla-mom}
\end{equation}
Moreover, since ${\Lambda}$ depends only on the overall coordinate transformations being
performed within the space ${\cal E}$, the difference on the left hand side of (\ref{totmomcons})
is also a ${4+2n-}$vector, and therefore transforms as
\begin{equation}
\Delta {\cal P}_{\!sys}'^A=\Lambda^A_{\,\;\;B}\;\Delta
{\cal P}_{\!sys}^B\,.\label{transla-differ}
\end{equation}
Thus, if the conservation law (\ref{totmomcons}) holds for one set of coordinates within
the space ${\cal E}$, then, according to (\ref{transla-differ}), it does so for {\it all}
coordinates related by the transformations (\ref{transla}). Consequently, the conservation
law (\ref{totmomcons}), once unpacked into its external, internal, and constituent parts as
\begin{equation}
0=\Delta {\boldsymbol{\cal P}}_{\!sys}=\left(\Delta E_{sys}/c\,,\;
\Delta {\bf p}_{sys}\,,\;\Delta {\bf P}_{sys}^{int}\right),\label{extmomcons}
\end{equation}
leads to the familiar conservation laws for energies and momenta: 
\begin{equation}
0=\Delta E_{sys}:=\sum_f E_f -\sum_i E_i\label{sumenegy}
\end{equation}
and
\begin{equation}
0=\Delta {\bf p}_{sys}:=\sum_f {\bf p}_f -\sum_i {\bf p}_i\,,\label{summom}
\end{equation}
where the indices ${f}$ and ${i}$ stand for the final and initial number of constituents of the
system. Thus in the present theory the energies and momenta remain as {\it additive} as in
special relativity. In other words, in the present theory not only are there no preferred class
of observers, but also the usual conservation laws of special relativity remain essentially
unchanged, contrary to expectation.

\subsection{The \textsl{raison d'\^etre} of time: causal inevitability of
becoming}\label{subsec:3.3}

With the physical structure of the generalized relativity in place, we are now in a position
to address the central concern of the present essay: namely, the {\it raison d'\^etre} of the
{\it tensed} time, as depicted in Fig. \ref{figure-2}. To this end, let us first note that the
causal structure embedded within our generalized relativity is profoundly unorthodox. One way to
appreciate this unorthodoxy is to recall the blurb for spacetime put forward by Minkowski in his
seminal address at Cologne, in 1908. ``Nobody has ever noticed a place except at a time, or
a time except at a place'', he ventured \cite{mink-quote}. But, surely, this famous quip of
Minkowski hardly captures the complete picture. Perhaps it is more accurate to say
something like: {\it Nobody has ever noticed a place except at a certain time while being in
a certain state, or noticed a time except at a certain place while being in a certain state, or
been in some state except at a certain time, and a certain place}. At any rate, this revised
statement is what better captures the notion of time afforded by our generalized theory
of relativity. For, as evident from the quadratic invariant (\ref{Minkowski-plus}), in
addition to space, time in our generalized theory is as much a state-dependent attribute as
states are time-dependent attributes, and as states of the world do happen and become, so does
time. Intuitively, this dynamic state of affairs can be summarized as follows:
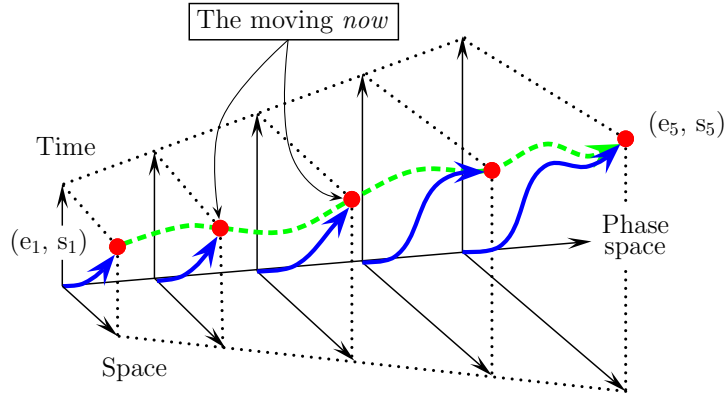
\begin{figure}
\hrule
\scalebox{0.7}{

\begin{pspicture}(0.0,1.0)(14.15,10.5)

\put(12.85,5.1){\Large{\rm Phase}}

\put(12.9,4.7){\Large{\rm space}}

\put(3.35,2.4){\Large{\rm Space}}

\put(2.1,6.6){\Large{\rm Time}}

\put(5.0,9.0){\psframebox{${{}^{}}$\Large{{\rm The moving} {\it now}}${{}^{}\,}$}}

\psline[arrowscale=2 3]{->}(2.6,4.1)(12.65,4.95)

\psline[linewidth=0.6mm, linestyle=dotted](2.6,6.03)(3.65,4.85)

\psline[linewidth=0.6mm, linestyle=dotted](3.65,3.15)(3.65,4.85)

\psline[linewidth=0.6mm, linestyle=dotted](4.35,6.65)(5.6,5.2)

\psline[linewidth=0.6mm, linestyle=dotted](5.65,2.92)(5.6,5.2)

\psline[linewidth=0.6mm, linestyle=dotted](6.3,7.38)(8.1,5.75)

\psline[linewidth=0.6mm, linestyle=dotted](8.1,2.74)(8.1,5.75)

\psline[linewidth=0.6mm, linestyle=dotted](8.3,8.13)(10.76,6.3)

\psline[linewidth=0.6mm, linestyle=dotted](10.76,2.42)(10.76,6.3)

\psline[linewidth=0.6mm, linestyle=dotted](10.2,8.8)(13.3,6.9)

\psline[linewidth=0.6mm, linestyle=dotted](13.3,2.1)(13.3,4.4)

\psline[linewidth=0.6mm, linestyle=dotted](13.3,5.7)(13.3,6.9)

\psline[linewidth=0.6mm, linestyle=dotted](2.6,6.03)(10.25,8.82)

\psline[linewidth=0.6mm, linestyle=dotted](3.65,3.15)(13.3,2.1)

\pscurve[linewidth=0.2mm,arrowscale=2 2]{->}(6.89,8.78)(5.6,7.2)(5.58,5.37)

\pscurve[linewidth=0.2mm,arrowscale=2 2]{->}(6.89,8.78)(6.95,6.6)(7.5,5.9)
(7.93,5.77)

\psline[arrowscale=2 3]{->}(4.35,4.25)(4.35,6.65)

\psline[arrowscale=2 3]{->}(6.3,4.38)(6.3,7.38)

\psline[arrowscale=2 3]{->}(8.3,4.55)(8.3,8.13)

\psline[arrowscale=2 3]{->}(10.2,4.75)(10.2,8.85)

\psline[arrowscale=2 3]{->}(4.35,4.25)(5.7,2.92)

\psline[arrowscale=2 3]{->}(6.3,4.38)(8.15,2.68)

\psline[arrowscale=2 3]{->}(8.3,4.55)(10.76,2.4)

\psline[arrowscale=2 3]{->}(10.2,4.75)(13.3,2.1)

\pscurve[linecolor=green, linewidth=0.9mm, linestyle=dashed, arrowscale=2]{->}(3.5,4.8)
(5,5.25)(5.6,5.2)(6.9,5.15)(8.1,5.75)(9.3,6.3)(10.73,6.31)(11.8,6.8)(12.6,6.57)
(13.00,6.72)(13.2,6.8)

\psline[arrowscale=2 3]{-}(2.6,4.1)(2.6,4.6)

\psline[arrowscale=2 3]{->}(2.6,5.25)(2.6,6.03)

\psdots[linecolor=red, fillcolor=red, dotstyle=o, dotscale=2.5](3.65,4.85)(5.6,5.2)(8.1,5.75)
(10.76,6.3)(13.3,6.9)

\pscurve[linecolor=blue,linewidth=0.8mm, arrowscale=2]{->}(2.61,4.1)(3.0,4.15)(3.6,4.7)

\psline[arrowscale=2 3]{->}(2.6,4.1)(3.65,3.15)

\pscurve[linecolor=blue,linewidth=0.8mm, arrowscale=2]{->}(4.4,4.2)(4.9,4.3)(5.55,5.05)

\pscurve[linecolor=blue,linewidth=0.8mm, arrowscale=2]{->}(6.3,4.38)(7,4.48)(7.8,5.3)(8.05,5.6)

\pscurve[linecolor=blue,linewidth=0.8mm, arrowscale=2]{->}(8.3,4.55)
(8.9,4.65)(9.8,6.1)(10.627,6.27)

\pscurve[linecolor=blue,linewidth=0.8mm, arrowscale=2]{->}(10.2,4.75)
(10.9,4.85)(11.8,6.4)(12.6,6.38)(13.2,6.8)

\put(1.6,4.8){\Large ${({\rm e}_1,\,{\rm s}_1)}$}

\put(13.72,7.05){\Large ${({\rm e}_5,\,{\rm s}_5)}$}

\end{pspicture}}
\hrule
\caption{Space-time-state diagram depicting the flow of time. The solid blue curves represent
growing timelike worldlines at five successive stages of growth, from ${{\rm s}_1}$ to
${{\rm s}_5}$, whereas the dashed green curve represents the growing overall worldline
from ${({\rm e}_1,\,{\rm s}_1)}$ to ${({\rm e}_5,\,{\rm s}_5)}$. The red dot represents the
necessarily moving present. In fact, the entire space-time-state structure is causally
necessitated to expand continuously.}
\label{figure-7}
\end{figure}
\begin{equation}
\begin{split}
{\rm x}&={\rm x}(t,\,{\rm y})\\
t&=t({\rm x},\,{\rm y})\\
{\rm y}&={\rm y}(t,\,{\rm x}),\label{intuit}
\end{split}
\end{equation}
where ${\rm y}$ is the phase space coordinate as before. In other words, place in the
present theory is regarded as a function of time and state; time is regarded as a function
of place and state; and state is regarded as a function of time and place. As we shall see,
it is this state-dependence of time that is essentially what mandates the causal necessity
for becoming in the present theory.

To appreciate this dynamic or tensed nature of time in the present theory, let us
return once again to our clock that is moving and evolving, say, from an event-state
${({\rm e}_1,\,{\rm s}_1)}$ to an event-state ${({\rm e}_5,\,{\rm s}_5)}$ in the combined
space ${\cal E}$, as depicted in the space-time-state\footnote{Here perhaps
``space-time-phase space diagram'' would be a much more accurate neologism, but it would be
even more mouthful than ``space-time-state diagram.''} diagram of Fig.
\ref{figure-7}. According to the line element (\ref{Minkowski-plus}), the proper duration
recorded by the clock would be given by
\begin{equation}
\Delta\tau=\int_{{\rm t}_1}^{{\rm t}_5}
\frac{1}{\gamma(v,\,\omega)}\;\;dt\,,\label{clock-duration}
\end{equation}
where ${\gamma(v,\,\omega)}$ is defined by (\ref{gengamm}). Now, assuming for simplicity
that the clock is not massless, we can represent its journey by the integral curve of a
timelike ${4+2n-}$velocity vector field ${V^A}$ on the space ${\cal E}$, defined by
\begin{equation}
V^A:={\cs lP}\frac{\,dz^A}{d\tau\;\,},
\end{equation}
such that its external components ${V^a}$ ${(a=0,1,2,3)}$ would trace out, for each possible
state ${s_i}$ of the clock, the familiar four dimensional timelike worldlines in the
corresponding Minkowski spacetime. In other words, the overall velocity vector field ${V^A}$
would give rise to the familiar timelike, future-directed, never vanishing, 4-velocity vector
field ${V^a}$, tangent to each of the external timelike worldlines. As a result, the ``length''
of the overall enveloping worldline would be given by the proper duration (\ref{clock-duration}),
whereas the ``length'' of the external worldline, for each ${s_i}$, would be given by the
Einsteinian proper duration
\begin{equation}
\Delta {\css {\tau}iE}=\int_{{\rm t}_1}^{{\rm t}_i}\frac{1}{\gamma(v)}\;dt\,,\label{yes-this}
\end{equation}
where ${\gamma(v)}$ is the usual Lorentz factor given by (\ref{us-gamma}). In Fig.
\ref{figure-7}, five of such external timelike
worldlines---one for each ${s_i}$ (i=1,2,3,4,5)---are depicted by the
blue curves with arrowheads going ``upwards'', and the overall enveloping worldline traced
out by ${V^A}$ is depicted by the dashed green curve going from the ``initial'' event-state
${({\rm e}_1,\,{\rm s}_1)}$ to the ``final'' event-state ${({\rm e}_5,\,{\rm s}_5)}$.

It is perhaps already clear from this picture that the external worldline of our clock is not
given all at once, stretched out till eternity, but grows continuously, along with each
temporally successive stage of the evolution of the clock, like a tendril on a wall. That is
to say, as anticipated in Fig. \ref{figure-2}, the future events along the external worldline
of the clock simply do not exist. Hence the ``now'' of the clock cannot even be said to be
preceding the future events, since, quite literally, there exists nothing to which it has the
relation of precedence \cite{Broad-1923}. Moreover, since the external Minkowski spacetime is
simply a congruence of non-intersecting timelike worldlines of idealized observers, according
to the present theory the entire sum total of existence must increase continuously
\cite{Broad-1923}. In fact, this continuous growth of existence turns out to be {\it causally
necessitated} in the present theory, and can be represented by a {\it Growth Vector Field}
quantifying the instantaneous directional rate of this growth:
\begin{equation}
U^a:={\hat V}^a\,\frac{d{\cs {\tau}E}}{d{\rm y}\;}\,,\label{47}
\end{equation}
where ${{\hat V}^a}$ is a unit vector field in the direction of the 4-velocity vector field
${V^a}$, ${d{\rm y}:=|d{\bf y}|}$ is the infinitesimal dimensionless phase space distance
between the two successive states of the clock discussed before, and ${d{\cs {\tau}E}}$ is the
infinitesimal Einsteinian proper duration defined by (\ref{Minkowski}). It is crucial to note
here that {\it in special relativity this Growth Vector Field would vanish identically
everywhere, whereas in our generalized theory it cannot possibly vanish anywhere}. This is
essentially because of the mutual dependence of place, time, and state in the present theory
we discussed earlier (cf. Eq. (\ref{intuit})). More technically, this is because the 4-velocity
vector ${V^a}$ of an observer in Minkowski spacetime, such as the one in Eq. (\ref{47}), can
never vanish, whereas the causality constraint (\ref{Minkowski-plus}) of the present theory
imposes the lower bound ${\cs tP}$ on the rate of change of Einsteinian proper duration with
respect to the phase space coordinates,
\begin{equation}
\frac{d{\cs {\tau}E}}{d{\rm y}\;}\geq {\cs tP}\,,\label{rateofnow}
\end{equation}
which, taken together,
{\it causally} necessitates the never-vanishing of the Growth Vector Filed ${U^a}$. Consequently,
the ``now'' of the clock (the red dot in Fig. \ref{figure-7}) moves in the future direction
along its external worldline, at the rate of no less than one Planck unit of time per Planck
unit of change in its physical state. And, along with the non-vanishing of the 4-velocity
vector field ${V^a}$, the lower bound ${\cs tP}$ on the growth rate of any external worldline
implies that not only do all such ``nows'' move, but they {\it cannot} not move---i.e., not only
does the sum total of existence increase, but it {\it cannot} not increase. To parody
Weyl quoted above, the objective world cannot simply {\it be}, it can only {\it happen}.

This conclusion can be further consolidated by realizing that in the present theory even the
overall enveloping worldline (the dashed green curve in Fig. \ref{figure-7}) cannot help but
grow non-relationally and continuously. This can be confirmed by first parallelling the above
analysis for the ${1+2n-}$dimensional internal space ${\cal O}$ instead of the external
spacetime ${\cal M}$, which amounts to slicing up the combined space ${\cal E}$ of Fig.
\ref{figure-7} along the spatial axis instead of the phase space axis, and then observing that
even the ``internal worldline'' (not shown in the figure) must necessarily grow progressively
further as time passes, at the rate given by the {\it internal} growth vector field
\begin{equation}
U^{\alpha}={\cs lP}\,{\hat V}^{\alpha}\frac{d{\cs tH}}{d{\rm x}\;}\,.\label{rateofthen}
\end{equation}
Here ${{\hat V}^{\alpha}}$ is a unit vector field in the direction of the ${1+2n-}$velocity
vector filed ${V^{\alpha}}$ corresponding to the internal part of the overall velocity vector
field ${V^A}$, ${d{\rm x}:=|d{\bf x}|}$ is the infinitesimal spatial distance between two slices,
and ${d{\cs tH}}$ is the infinitesimal internal proper duration defined by Eq. (\ref{a-metric}).
Once again, it is easy to see that
the causality condition (\ref{Minkowski-plus}) gives rise to the lower bound
\begin{equation}
{\cs lP}\,\frac{d{\cs tH}}{d{\rm x}\;}\geq {\cs tP}\,.\label{sett}
\end{equation}
Thus, ``now'' of the clock necessarily moves in the future direction also along its internal
worldline within the internal space ${\cal O}$. As a result, even the overall
worldline---namely, the dashed green curve in Fig. \ref{figure-7}---can be easily shown to be
growing non-relationally and continuously. Indeed, using Eqs. (\ref{47}) to (\ref{sett}),
an elementary geometrical analysis \cite{Christian-2004} shows that the instantaneous
directional rate of this growth is given by the {\it overall} growth vector field
\begin{equation}
U^A=\left({\hat V}^a\,\frac{d{\cs tE}}{d{\rm y}\;}\,,\;
{\cs lP}{\hat V}^{\mu}\,\frac{d{\cs tH}}{d{\rm x}\;}\right),\label{totalrate}
\end{equation}
whose magnitude also remains bounded from below by the Planck time ${\cs tP}$:
\begin{equation}
\sqrt{-\xi_{AB}U^AU^B}\;\geq\;{\cs tP}\,.\label{U-bound}
\end{equation}
Thus, in the present theory, not only are the external events in ${\cal E}$ not all laid out once
and for all, for all eternity, but there does not remain even an overall ${4+2n-}$dimensional
``block'' that could be used to support a ``block'' view of the universe. In fact, the causal
necessity of the lower bound (\ref{U-bound}) on the magnitude of the overall growth vector
field ${U^A}$---which follows from the causality constraint (\ref{Minkowski-plus})---exhibits
that in the present theory the sum total of existence itself is causally necessitated to
increase continuously. That is to say, the very structure of the present theory {\it causally
necessitates} the universe to be purely {\it Heraclitean}, in the sense discussed in the
Introduction.

\section{Prospects for the experimental metaphysics of time}\label{sec:4}

As alluded to in the Introduction, any empirical confrontation of the above generalized
relativity with special relativity would amount to a step towards what may be called the
experimental metaphysics of time. However, since the generalized theory is deeply rooted in
the Planck regime, any attempt to experimentally discriminate it from special relativity
immediately encounters a formidable practical difficulty. To appreciate this difficulty, 
consider the following series expansion of expression (\ref{t}) for the generalized proper
time, up to second order in the Planck time:
\begin{equation}
\Delta\tau=\Delta t\,\sqrt{1-\frac{v^2}{c^2}} \,-\,
\frac{1}{2}\,\frac{\,\css t2P}{\Delta t}\left(1-\frac{v^2}{c^2}\right)^{\!-\frac{3}{2}}\!+
\dots\label{exp-t}
\end{equation}
The first term on the right hand side of this expansion is, of course, the familiar special
relativistic term. The difficulty arises in the second term, i.e. in the first largest
correction term to the special relativistic time dilation effect, since this term is
modulated by the {\it square} of the Planck time, which in ordinary units amounts to some
${10^{-87}}$ ${{sec}^2}$. Clearly, the precision required to directly verify such a
miniscule correction to the special relativistic prediction is well beyond
the scope of any foreseeable precision technology.

Fortunately, in recent years an observational possibility has emerged that might save the
day for the experimental metaphysics of time. The central idea that has emerged during the
past decade within the context of quantum gravity is to counter the possible Planck scale
suppression of physical effects by appealing to ultrahigh energy particles cascading
the earth that are produced at cosmological distances. One strategy along this line is to
observe oscillating flavor ratios of ultrahigh energy cosmic neutrinos to detect possible
deviations in the energy-momentum relations predicted by special relativity
\cite{Christian-2005}. Let us briefly look at this strategy, as it is applied to our
generalized theory of relativity (further details can be found in Refs. \cite{Christian-PRL}
and \cite{Christian-2005}; as in these references, from now on we shall be using the Planck
units: ${\hbar=c=G=1}$).

\subsection{Testing Heraclitean relativity using cosmic neutrinos}\label{subsec:4.1}

The remarkable phenomena of neutrino oscillations are due to the fact that neutrinos of definite
flavor states ${\ket{\nu_{\alpha}}}$, ${\alpha=e,\mu,}$ or ${\tau}$, are {\it not} particles of
definite mass states ${\ket{{\nu_j}}}$, ${j=1,2,}$ or ${3}$, but are superpositions of the
definite mass states. As a neutrino of definite flavor state propagates through vacuum for a
long enough laboratory time, its heavier mass states lag behind the lighter ones, and the
neutrino transforms itself into an altogether different flavor state. The probability for this
``oscillation'' from a given flavor state, say ${\ket{\nu_{\alpha}(0)}}$, to another
flavor state, say ${\ket{\nu_{\beta}(t)}}$, is famously given by
\begin{equation}
P_{\alpha\beta}(E,\,L) =\delta_{\alpha\beta}-\!\sum_{j\not=k}U^*_{\alpha j}U_{\alpha k}
U_{\beta j}U^*_{\beta k}\!\left[1-e^{-i(\Delta m^2_{jk}/2E)L}\right]\!. \label{transi-prob}
\end{equation}
Here ${\Delta m_{jk}^2}$ ${\equiv}$ ${m_k^2-m_j^2}$ ${> 0}$ is the difference in the squares
of the two neutrino masses, ${U}$ is the time-independent leptonic mixing matrix, and ${E}$
and ${L}$ are, respectively, the energy and distance of propagation of the neutrinos. It is
clear from this transition probability that the experimental observability of the flavor
oscillations is dependent on the quantum phase
\begin{equation}
\Phi:=2\pi\,\frac{L\,}{L_O}\,,\label{quan-phase}
\end{equation}
where
\begin{equation}
L_O :=\frac{2\,\pi}{\Delta p}=\frac{4\pi E}{\,\Delta m_{jk}^2}\label{osci-length}
\end{equation}
is the energy-dependent oscillation length. Thus, changes in neutrino flavors would be
observable whenever the propagation distance ${L}$ is of the order of the oscillation length
${L_O}$. However, in definition (\ref{osci-length}) the difference in momenta,
${{\Delta p}\equiv p_j-p_k}$, was obtained by using the special relativistic relation
\begin{equation}
p_j=\sqrt{E^2-m_j^2}\,\approx E-\frac{\,m^2_j\,}{2E\,}.\label{momen}
\end{equation}
In the present theory this relation between energies and momenta is, of course, generalized,
and given by (\ref{exact-disp}), replacing the above approximation by
\begin{equation}
p_j\approx E-\frac{m_j^2}{2E}+\frac{E^2}{\css m2P}\,m_j\label{mod-mom}
\end{equation}
up to the second order, with ${\cs mP}$ being the Planck mass. The
corresponding modified oscillation length analogous to (\ref{osci-length}) is then given by
\begin{equation}
L^{'}_O :=\frac{2\,\pi}{\Delta p}=\frac{2\,\pi}{\frac{1}{2E}\,\Delta m_{jk}^2\,-\,
\frac{E^2}{\,\css m2P}\,\Delta m_{jk}}\,,\label{r-osci-r}
\end{equation}
where ${\Delta m_{jk}^2 \equiv m^2_k-m^2_j}$ as before, and ${\Delta m_{jk} \equiv m_k-m_j > 0}$.
Consequently, according to our generalized relativity the transition probability
(\ref{transi-prob}) would be quite different in general, as a function of ${E}$ and
${L}$, from how it is according to special relativity. And despite the {\it quadratic} Planck
energy suppression of the correction to the oscillation length, this difference would be
observable for neutrinos of sufficiently high energies and long propagation
distances. Indeed, it can be easily shown \cite{Christian-2005} that the relation 
\begin{equation}
L\sim\frac{\,\pi\,\css m4P}{E^5}\label{L-right}
\end{equation}
is the necessary constraint between the neutrino energy ${E}$ and the propagation distance ${L}$
for the observability of possible deviations from the standard flavor oscillations. For instance,
it can be readily calculated from this constraint that the Planck scale deviations in the
oscillation length predicted by our generalized relativity would be either observable, or can
be ruled out, for neutrinos of energy ${E\sim 10^{17}}$ eV, provided that they have originated
from a cosmic source located at some ${10^{5}}$ light-years away from a terrestrial detector.
The practical means by which this can be achieved in the foreseeable future have been discussed
in some detail in the Refs. \cite{Christian-PRL} and \cite{Christian-2005} cited above.

\subsection{Testing Heraclitean relativity using ${\gamma}$-ray binary pulsars}\label{subsec:4.2}

The previous method of confronting the generalized theory of relativity with special relativity
is clearly phenomenological. Fortunately, a much more direct test of the generalized theory may
be possible, thanks to the precise deviations it predicts from the special relativistic Doppler
shifts \cite{Christian-2004}:
\begin{equation}
\frac{\,E'}{E\;}=\frac{\,\varepsilon'\left[\,\varepsilon'-\frac{v}{c}\cos\phi\,\right]}
{\sqrt{(\varepsilon')^2-\frac{v^2}{c^2}\,}\;}\,,\label{newpho}
\end{equation}
with
\begin{equation}
\varepsilon':=\sqrt{1-\frac{E^2}{\css E2P}\!
\left(1-\frac{E'}{E\;}\right)^{\!2}}\,.\label{(1')}
\end{equation}
Here ${v}$ is the relative speed  of a receiver receding from a photon source, ${E}$ and
${E'}$, respectively, are the energy of the photon and that observed by the receiver,
and ${\phi}$ is the angle between the velocity of the receiver and the photon momentum.
Note that ${\varepsilon'}$ here clearly reduces to unity for ${E'-E\ll{\cs EP}}$, thus
reducing the generalized expression (\ref{newpho}) to the familiar {\it linear} relation
for Doppler shifts predicted by special relativity.

Even without solving the relation (\ref{newpho}) for ${E'}$ in terms of ${E}$, it is not
difficult to see that, since ${\varepsilon'< 1}$, at sufficiently high energies any red-shifted
photons would be somewhat more red-shifted according to (\ref{newpho}) than predicted by special
relativity. But one can do better than that. A Maclaurin expansion of the right hand side of
(\ref{newpho}) around the value ${E/{\cs EP}=0}$, after keeping terms only up to the second
order in the ratio ${E/{\cs EP}}$, gives
\begin{equation}
\frac{\,E'}{E\,}\approx\frac{1-\frac{v}{c}\cos\phi}{\sqrt{1-\frac{v^2}{c^2}}\;\,}
+\frac{1}{2}\frac{E^2}{\css E2P}\!\left[\frac{1-\frac{v}{c}\cos\phi}{\left(1-\frac{v^2}{c^2}
\right)^{3/2}}-\frac{2-\frac{v}{c}\cos\phi}{\left(1-\frac{v^2}{c^2}
\right)^{1/2}}\right]\!\!\left(\!1-\frac{\,E'}{E\,}\right)^2\!\!\!+\dots\label{series-a}
\end{equation}
This truncation is an excellent approximation to (\ref{newpho}).
The quadratic equation (\ref{series-a}) can now be solved for the desired ratio ${E'/E}$, and
then the physical root once again expanded, now in the powers of ${v/c}$. In what results if we
again keep terms only up to the second order in the ratios ${E/{\cs EP}}$ and ${v/c}$, then,
after some tedious but straightforward algebra, we arrive at
\begin{equation}
\frac{\,E'}{E\,}\!\approx
1-\frac{v}{c}\cos\phi+\frac{1}{2}\left[1-\frac{E^2}{\css E2P}\cos^2\!\phi\right]
\frac{v^2}{c^2}\pm\dots,\label{prac-inter}
\end{equation}
which, in the limit ${E\ll{\cs EP}}$, reduces to the special relativistic result
\begin{equation}
\frac{\,E'}{E\,}\!\approx\!
1-\frac{v}{c}\cos\phi+\frac{1}{2}\frac{v^2}{c^2}\pm\dots.\label{newpa-rela}
\end{equation}

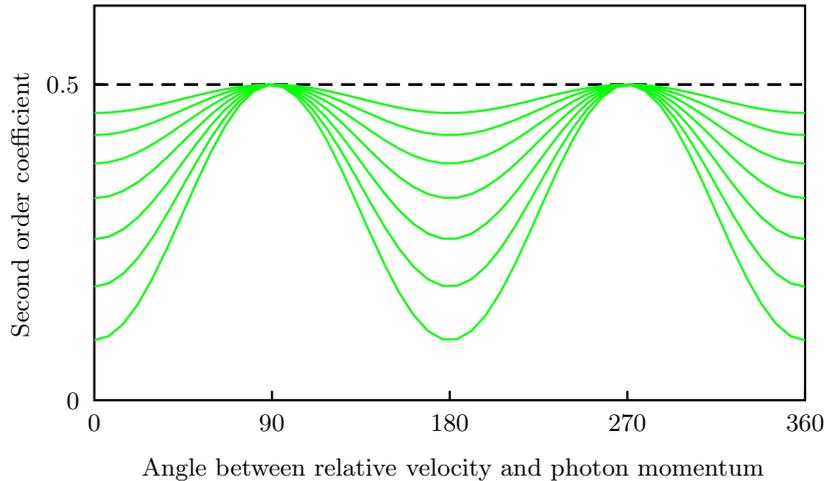
\begin{figure}
\hrule
\scalebox{1.05}{
\begin{pspicture}(-1.3,-1.5)(8.87,5.5)
\psset{xunit=.25mm,yunit=4cm}
\psaxes[axesstyle=frame,tickstyle=top,dx=90\psxunit,Dx=90,dy=1\psyunit,Dy=0.5](0,0)(360,1.25)
\psplot[linewidth=.35mm,linecolor=black,linestyle=dashed]{0.0}{255}{x 1.0}
\psplot[linewidth=.27mm,linecolor=green]{0.0}{360}{x dup cos exch cos mul 0.81 mul neg 1 add}
\psplot[linewidth=.27mm,linecolor=green]{0.0}{360}{x dup cos exch cos mul 0.64 mul neg 1 add}
\psplot[linewidth=.27mm,linecolor=green]{0.0}{360}{x dup cos exch cos mul 0.49 mul neg 1 add}
\psplot[linewidth=.27mm,linecolor=green]{0.0}{360}{x dup cos exch cos mul 0.36 mul neg 1 add}
\psplot[linewidth=.27mm,linecolor=green]{0.0}{360}{x dup cos exch cos mul 0.25 mul neg 1 add}
\psplot[linewidth=.27mm,linecolor=green]{0.0}{360}{x dup cos exch cos mul 0.16 mul neg 1 add}
\psplot[linewidth=.27mm,linecolor=green]{0.0}{360}{x dup cos exch cos mul 0.09 mul neg 1 add}
\put(0.6,-0.95){Angle between relative velocity and photon momentum}
\put(-1.05,0.8){\rotatebox{90}{Second order coefficient}}
\end{pspicture}}
\hrule
\caption{The energy-dependent signatures of Heraclitean relativity. The green curves are based
on the predictions of the present theory, for ${E/{\cs EP}=0.3}$ to ${0.9}$ in the descending
order, whereas the dashed black line is the prediction of special relativity.}
\label{figure-8}
\end{figure}

Comparing (\ref{prac-inter}) and (\ref{newpa-rela}) we see that up to the first order in ${v/c}$
there is no difference between the special relativistic result and that of the present theory.
The first deviation between the two theories occur in the second-order coefficient, precisely
where special relativity differs also from the classical theory. What is more, this second-order
deviation depends non-trivially on the angle between the relative velocity and photon momentum.
For instance, up to the second order both red-shifts (${\phi=0}$) and blue-shifts (${\phi=\pi}$)
predicted by (\ref{prac-inter}) significantly differ from those predicted by special relativity.
In particular, the red-shifts are now somewhat more red-shifted, whereas the blue-shifts are
somewhat less blue-shifted. On the other hand, the transverse red-shifts (${\phi=\pi/2}$ or
${\phi=3\pi/2}$) remain identical to those predicted by special relativity. As a result, even
for the photon energy approaching the Planck energy an Ives-Stilwell type classic experiment
\cite{Ives-Stilwell} would not be able to distinguish the predictions of the present theory
from those of special relativity. The complete angular distribution of the second-order
coefficient predicted by the two theories, along with its energy dependence, is displayed in
Fig. \ref{figure-8}.

In spite of this rather non-trivial angular dependence of Doppler shifts, in practice, due to
the quadratic suppression by Planck energy, distinguishing the expansion (\ref{prac-inter}) from
its special relativistic counterpart (\ref{newpa-rela}) would be a formidable task. The maximum
laboratory energy available to us is of the order of ${10^{12}}$ eV, yielding
${E^2/{\css E2P}\sim 10^{-32}}$. This represents a correction of one part in ${10^{32}}$ from
(\ref{newpa-rela}), demanding a phenomenal sensitivity of detection well beyond the means
of any foreseeable precision technology. However, an extraterrestrial source, such as an
extreme energy ${\gamma}$-ray binary pulsar, may turn out to be accessible for distinguishing
the second order Doppler shifts predicted by the two theories. It is well known that binary
pulsars not only exhibit Doppler shifts, but the second-order shifts resulting from the
periodic motion of such a pulsar about its companion can be isolated, say, from the first order
shifts, because they depend on the {\it square} of the relative velocity, which varies as the
pulsar moves along its two-body elliptical orbit \cite{Will-1993}. Due to these Doppler shifts,
the rate at which the pulses are observed on Earth reduces slightly when the pulsar is receding
away from the Earth, compared to when it is approaching towards it.
As a result, the period, its variations, and other orbital characteristics of the pulsar, as
they are determined on Earth, crucially depend on these Doppler shifts. In practice, the
parameter relevant in the arrival-time analysis of the pulses received on Earth turns out to be
a non-trivial function of the gravitational red-shift, the masses of the two binary stars, and
other Keplerian parameters of their orbits, and is variously referred to as the
red-shift-Doppler parameter or the time dilation parameter \cite{Will-1993}. For a pulsar that
is also following a periastron precession similar to the perihelion advance of Mercury, it can
be determined with excellent precision.

The arrival-time analysis of the pulses begins by considering the time of emission of the
${N^{th}}$ pulse, which is given by
\begin{equation}
N={\cs NO} +\nu\tau+\frac{1}{2}{\dot\nu}\tau^2+\frac{1}{6}{\ddot\nu}\tau^3+\dots\,,
\end{equation}
where ${\cs NO}$ is an arbitrary integer, ${\tau}$ is the proper time measured by a clock in
an inertial frame on the surface of the pulsar, and ${\nu}$ is the rotation frequency of the
pulsar, with ${{\dot\nu}\equiv d{\nu}/d{\tau}|_{\tau=0}}$ and
${{\ddot\nu}\equiv d^2{\nu}/d{\tau}^2|_{\tau=0}}$. The proper time ${\tau}$ is related to the
coordinate time ${t}$ by
\begin{equation}
d\tau\!=dt\!\left[1-\frac{\alpha_2^* m_2}{r}-\frac{1}{2}\frac{v_1^2}{c^2}+...\,\right]\!,
\label{verymuch}
\end{equation}
where the first correction term represents the gravitational red-shift due to the field of the
companion, and the second correction term represents the above mentioned second-order Doppler
shift due to the orbital motion of the pulsar itself. The time of arrival of the pulses on Earth
differs from the coordinate time ${t}$ taken by the signal to travel from the pulsar to the
barycenter of the solar system, due to the geometrical intricacies of the pulsar binary and
the solar system \cite{Will-1993}. More relevantly for our purposes, the time of arrival of the
pulses is directly affected by the second-order Doppler shift appearing in Eq. (\ref{verymuch}),
which thereby affects the observed orbital parameters of the pulsar.

Now, returning to our Heraclitean generalization of relativity, it is not difficult to see that
the generalized Doppler shift expression (\ref{prac-inter}) immediately gives the following
generalization of the infinitesimal proper time  (\ref{verymuch}):
\begin{equation}
d\tau=dt\!\left[1-\frac{\alpha_2^* m_2}{r}-\frac{1}{2}\left\{\!1-
\frac{E^2}{\css E2P}\cos^2\phi\!\right\}\frac{v_1^2}{c^2}+...\,\right]\!.
\label{correct}
\end{equation}
Thus, in our generalized theory the second-order Doppler shift acquires an {\it energy-dependent}
modification. The question then is: At what radiation energy this nontrivial modification will
begin to affect the observable parameters of the pulsar? The most famous pulsar, namely PSR
B1913+16, which has been monitored for three decades with exquisite accumulation of timing data,
is a radio pulsar, and hence for it the energy-dependent modification predicted in
(\ref{correct}) is utterly negligible, thanks to the quadratic suppression by the Planck energy.
However, for a ${\gamma}$-ray pulsar with sufficiently high radiation energy the modification
predicted in (\ref{correct}) should have an impact on its observable parameters, such
as the orbital period and its temporal variations.

The overall precision in the timing of the pulses from PSR B1913+16, and consequently in the
determination of its orbital period, is famously better than one part in ${10^{14}}$
\cite{Will-2006}. Indeed, the monitoring of the decaying orbit of PSR B1913+16 constitutes one of
the most stringent tests of general relativity to date. It is therefore not inconceivable that
similar careful observations of a suitable ${\gamma}$-ray pulsar may be able to distinguish the
predictions of the present theory from those of special relativity. Unfortunately, the highest
energy of radiation from a pulsar known to date happens to be no greater than ${10^{13}}$ eV,
giving the discriminating ratio ${E^2/{\css E2P}}$ to be of the order of ${10^{-30}}$, which is
only two orders of magnitude improvement over a possible terrestrial scenario. On the other hand,
the ${\gamma}$-rays emitted by a binary pulsar would have to be of energies exceeding ${10^{21}}$
eV for them to have desired observable consequences, comparable to those of PSR B1913+16.
Moreover, the desired pulsar have to be located sufficiently nearby, since above the ${10^{13}}$
eV threshold ${\gamma}$-rays are expected to attenuate severely through pair-production if they
are forced to pass through the cosmic infrared background before reaching the Earth. It is not
inconceivable, however, that a suitable binary pulsar emitting radiation of energies exceeding
${10^{21}}$ eV is found in the near future, allowing experimental discrimination of our
generalized relativity from special relativity.

\section{Concluding remarks}\label{sec:5}

One of the perennial problems in natural philosophy is the problem of change; namely, {\it How
is change possible?} Over the centuries, this problem has fostered two diametrically opposing
views of time and becoming. While these two views tend to agree that time presupposes change,
and that genuine change requires becoming, one of them actually denies the reality of change
and time, by rejecting becoming as a ``stubbornly persistent illusion'' \cite{Einstein-Besso}.
The other view, by contrast, accepts the reality of change and time, by embracing becoming as
a {\it bona fide} attribute of the world. Since the days of Aristotle within physics we have
been rather successful in explaining {\it how} the changes occur in the world, but seem to
remain oblivious to the deeper question of {\it why} do they occur at all. The situation has
been aggravated by the advent of Einstein's theories of space and time, since in these theories
there is no room to {\it structurally} accommodate the distinction between the past and the
future---a prerequisite for the genuine onset of change. By contrast, the causal structure of
the Heraclitean relativity discussed above not only naturally distinguishes the past form the
future by causally necessitating becoming, but also {\it forbids} inaction altogether, thereby
providing an answer to the deeper question of change. Moreover, since it is not impossible
to experimentally distinguish the Heraclitean relativity from special relativity, and since the
ontology underlying only the latter of these two relativities is prone to a block universe
interpretation, the enterprise of {\it experimental metaphysics of time} becomes feasible now,
for the first time, within a relativistic context. At the very least, such an enterprise should
help us decide whether time is best understood relationally, or non-relationally.

\section*{Acknowledgments}

I would like to thank Huw Price and Abner Shimony for their comments on Ref.
\cite{Christian-2004}, of which this essay is an apologia. I would also like to thank
Lucien Hardy, Lee Smolin, Antony Valentini, and other members of the Foundations of Physics
group at the Perimeter Institute for their hospitality and support.

\end{document}